\documentclass[manuscript,screen,sigconf]{acmart}

\AtBeginDocument{%
  \providecommand\BibTeX{{%
    \normalfont B\kern-0.5em{\scshape i\kern-0.25em b}\kern-0.8em\TeX}}}

\copyrightyear{2022}
\acmYear{2022}
\setcopyright{acmlicensed}\acmConference[PETRA '22]{The15th International Conference on PErvasive Technologies Related to Assistive Environments}{June 29-July 1, 2022}{Corfu, Greece}
\acmBooktitle{The15th International Conference on PErvasive Technologies Related to Assistive Environments (PETRA '22), June 29-July 1, 2022, Corfu, Greece}
\acmPrice{15.00}
\acmDOI{10.1145/3529190.3529214}
\acmISBN{978-1-4503-9631-8/22/06}

\begin{document}


\title{Diabetic foot ulcers monitoring by employing super resolution and noise reduction deep learning techniques}


\author{Agapi Davradou}
\affiliation{%
  \institution{National Technical University of Athens}
  \country{Greece}}
\email{adavradou@mail.ntua.gr}
\author{Eftychios Protopapadakis}
\affiliation{%
  \institution{National Technical University of Athens}
  \country{Greece}}
\email{eftprot@mail.ntua.gr}
\author{Maria Kaselimi}
\affiliation{%
  \institution{National Technical University of Athens}
  \country{Greece}}
\email{mkaselimi@mail.ntua.gr}
\author{Anastasios Doulamis}
\affiliation{%
  \institution{National Technical University of Athens}
  \country{Greece}}
\email{adoulam@cs.ntua.gr}
\author{Nikolaos Doulamis}
\affiliation{%
  \institution{National Technical University of Athens}
  \country{Greece}}
\email{ndoulam@cs.ntua.gr}

\renewcommand{\shortauthors}{Davradou et al.}

\begin{abstract}

Diabetic foot ulcers (DFUs) constitute a serious complication for people with diabetes. The care of DFU patients can be substantially improved through self-management, in order to achieve early-diagnosis, ulcer prevention, and complications management in existing ulcers. In this paper, we investigate two categories of image-to-image translation techniques (ItITT), which will support decision making and monitoring of diabetic foot ulcers: noise reduction and super-resolution. In the former case, we investigated the capabilities on noise removal, for convolutional neural network stacked-autoencoders (CNN-SAE). CNN-SAE was tested on RGB images, induced with Gaussian noise. The latter scenario involves the deployment of four deep learning super-resolution models. The performance of all models, for both scenarios, was evaluated in terms of execution time and perceived quality. Results indicate that applied techniques consist a viable and easy to implement alternative that should be used by any system designed for DFU monitoring.

\end{abstract}



\maketitle

\section{Introduction}

The "prevention is better than cure" principle applies aptly in diabetic foot ulcer (DFU) and can be achieved by motivating patients towards self-monitoring. People involved in self-management can help prevent/postpone the appearance of an ulcer, by detecting the corresponding signs and symptoms early on. Additionally, monitoring of existing ulcers is advised to prevent complications or recurrent ulceration \cite{armstrong2017diabetic}. During DFU monitoring, there are various signs and symptoms that should be taken under consideration including: skin color change (redness), skin temperature change, foot pressure induced injury (damage to the skin and/or underlying soft tissue), pain, swelling, or odor. 

Today, most of these DFU indication signs can be captured and, consequently, monitored, using various optical sensors \cite{doulamis2021non}, such as those integrated in mobile devices \cite{goyal2018robust} . It is intriguing the fact that RGB and thermal sensors can support DFU monitoring. Some major advantages of these type of sensors involve the, relatively, low acquisition costs, compact structure, and easy integration to portable devices. Typically, the raw data provided by the sensors are fed to complex AI tools, which serve as decision making mechanisms. There are multiple studies advocating that AI tools, coupled with optical sensors, can provide extremely useful decision support mechanisms. Such mechanisms can assist both physicians and patients to prevent undesirable situations \cite{tulloch2020machine}.

In this paper, we focus on the AI tools for image-to-image translation techniques (ItITT) \cite{kaji2019overview}. This scenario considers the steps prior to any decision making. The main research question focuses on how ItITT may support decision making and monitoring of DFU, given a set of images, provided by devices. Two cases of ItITT are considered: a) noise reduction and b) super-resolution (SR). Common reasons for low-quality images in wound documentation may appeared due to poor focus, motion blur, occlusion, inadequate lighting, and backlight, or due to time constraints associated with treatment and documentation, even when the image capturing is performed by trained personnel \cite{yap2021deep}. Improving the quality of a medical image is crucial to support the decision making and diagnosis \cite{hind2012noise}. 

The remainder of the paper is structured as follows: section \ref{section:rel_work} presents a brief background on noise removal using autoencoders and super-resolution techniques in medical imagery data. Section \ref{section:mod_arch} describes the proposed deep learning tool to improve the quality in medical images for diabetic foot ulcer monitoring. In Section \ref{section:exp_set}, an extensive experimental evaluation of the discussed methods is provided, while Section \ref{section:concl} closes the paper with a summary of findings.

\section{Related work}
\label{section:rel_work}

Applying deep learning techniques for ulcer detection \cite{yap2021deep} and segmentation \cite{goyal2017fully}, as well as classification of infection and ischaemia \cite{goyal2020recognition}, \cite{yap2021analysis} of diabetic foot ulcers, has shown great potential.
Yet little emphasis was given for preprocessing/image enhancement steps. Several attempts have been made in order to tackle the image denoising problem using deep learning (DL) and autoencoders have been one of the main applied techniques \cite{voulodimos2018deep}. In \cite{vincent2008extracting} denoising autoencoders were presented, in order to learn features from noisy images, while \cite{burger2012image} used stacked denoising autoencoders to reconstruct clean images from noisy images by exploiting the encoding layer of the multilayer perceptron (MLP). In \cite{xie2012image} Gaussian noise removal was achieved, by applying stacked sparse denoising auto-encoders and \cite{gondara2016medical} showed that denoising autoencoders constructed using convolutional layers can be efficiently used to denoise medical images. In \cite{liu2018low} stacked sparse denoising autoencoders were used to reduce the noise and improve the overall imaging quality of low-dose computed tomography (CT) images and \cite{ghosh2019sdca} applied stack  denoising  convolutional  autoencoder for retinal image denoising.

\par Conventional SR methods learn the dictionaries \cite{timofte2013anchored} or manifolds \cite{bevilacqua2012low} for modeling of the patch space. On the other hand, deep neural network (DNN)-based methods learn an end-to-end mapping between low- and high-resolution images, thus implicitly achieving dictionaries or mapping functions for patch space by hidden layers of DNN. With SRCNN (Super-Resolution Convolutional Neural Network), low-resolution input images are first upscaled to the desired size by use of bicubic interpolation and are then fed to an encoder-decoder network; thus, end-to-end mapping between the bicubic upscaled version of a low-resolution image and a ground truth high-resolution image is learned. 

The SRCNN has been applied to mammography images \cite{umehara2017super} and chest CT images \cite{umehara2018application}. SRGAN (Super-resolution using a generative adversarial network) \cite{ledig2017photo} can be thought of as a generative adversarial network (GAN)-fortified version of SRCNN. An encoder-decoder network with more upsamling layers than downsampling layers is trained to recover detailed textures from heavily downsampled images, and a discriminator is trained to differentiate between the super-resolved images and original high-quality images. SRGAN has been applied to the generation of high-resolution brain magnetic resonance imaging (MRI) images from low-resolution images, and 3D convolution has been adopted for exploiting volumetric information \cite{sanchez2018brain}.

\section{Models' architecture}
\label{section:mod_arch}

\subsection{Noise Removal Tool}

In this scenario, the capabilities on noise removal were investigated, for a convolutional neural network stacked-autoencoders (CNN-SAE) over RGB images, induced with Gaussian noise. The concept imposes the generation of random noise over multiple images, the training of a CNN-SAE model and the application over multiple test sets (i.e., previously unseen images). The use of high variance values have been intentionally selected, in order for the noise to stress test the model capabilities.

\begin{figure}[h]
  \centering
  \includegraphics[width=0.95\linewidth]{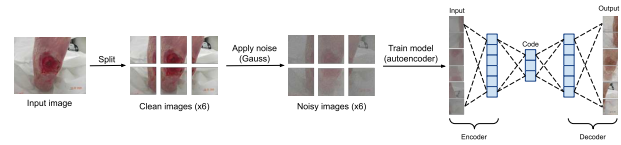}
  \caption{Training phase illustration for the deep learning model.}
  \Description{Training phase illustration for the deep learning model.}
  \label{fig:denoising_train}
\end{figure}

Figure \ref{fig:denoising_train} demonstrates the steps during the training process. Initially, an RGB image, denoted as $I$, is split in smaller patches $I^{(k)}$, of size 256×256 pixels each. Then, Gaussian noise of \((\mu,\sigma)=(0,0.3)\) is applied to each of the patches, creating a new (noisy) image patch   $\tilde{I}^{(k)}$. Then, these pairs  ($I^{(k)}$,$\tilde{I}^{(k)}$) are gathered and serve as a training set for the deep learning (DL) model.

\begin{figure}[h]
  \centering
  \includegraphics[width=0.8\linewidth]{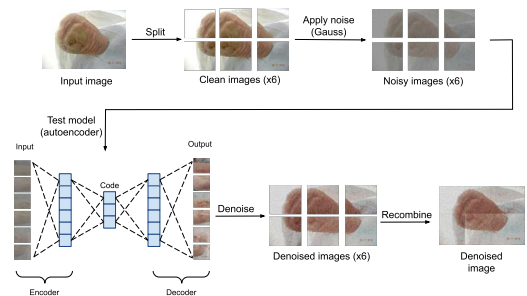}
  \caption{Testing phase illustration for the DL model.}
  \Description{Testing phase illustration for the DL model.}
  \label{fig:denoising_test}
\end{figure}

Figure \ref{fig:denoising_test}  demonstrates the testing phase for the model. The adopted approach allows for the fast implementation of the proposed CNN-SAE scheme over any type of image. The stitching part, i.e., merging the denoised patches back to a whole image can be further refined to mitigate the lining patterns.

\subsection{Super-resolution Tool}

\begin{figure}[h]
  \centering
  \includegraphics[width=0.95\linewidth]{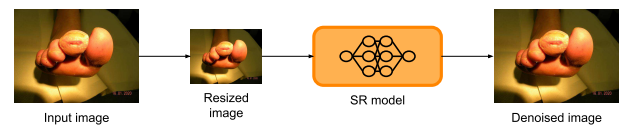}
  \caption{The implemented deep learning super-resolution evaluation scheme.}
  \Description{The implemented deep learning super-resolution scheme.}
  \label{fig:SR_model}
\end{figure}

Four deep learning techniques were selected and investigated to tackle the super-resolution task: SRGAN \cite{ledig2017photo}, EDSR \cite{lim2017enhanced}, ESRGAN \cite{wang2018esrgan} and ISR \cite{cardinale2018isr} . The pipeline for the utilization of the super-resolution methods is illustrated in Figure \ref{fig:SR_model}.

As mentioned in section \ref{section:rel_work}, the SRGAN is a GAN for image super-resolution for which a deep residual network (ResNet) \cite{he2016deep} with skip-connection was employed. The authors proposed a new perceptual loss function, which consists of an adversarial loss and a content loss. The adversarial loss uses a discriminator network, that is trained to differentiate between the super-resolved images and original photo-realistic images, in order to push the solution to the natural image manifold. The content loss was motivated by perceptual similarity instead of similarity in pixel space and is calculated on high-level feature maps of the VGG network \cite{johnson2016perceptual}, which are more invariant to changes in pixel space \cite{li2016combining}. 

The ESRGAN is an enhanced version of SRGAN in three aspects. First, the network structure was improved by introducing the residual-in-residual dense block (RDDB), without batch normalization (BN) \cite{ioffe2015batch} layers as in \cite{lim2017enhanced}.  Second the discriminator was improved using relativistic average GAN (RaGAN) \cite{jolicoeur2018relativistic}, which learns to judge the relative realness of an image, instead of the absolute value, whether it is real or not. Finally, the perceptual loss was also improved by using the VGG features before activation instead of after activation.

The EDSR is an enhanced deep super-resolution network based on the SRResNet \cite{ledig2017photo} architecture. The model is optimized by removing unnecessary modules in conventional residual networks. In addition, scale-independent information is utilized during the training phase, by training high-scale models from pre-trained low-scale models. 

Finally, the ISR model performs image super-resolution by combining implementations of different residual dense networks. More specifically, the model utilizes the SRGAN custom discriminator network, a multi-output version of the VGG19 network for deep features extraction for the perceptual loss, the super-scaling residual in residual dense network of ESRGAN and finally, the super-scaling residual dense network described in \cite{zhang2018residual}.

\section{Experimental setup}
\label{section:exp_set}

\subsection{Dataset description}
In this work the DFUC 2020 dataset, from the Diabetic Foot Ulcers Grand Challenge 2020 (DFUC 2020), is utilized \cite{cassidy2021dfuc}, containing clinically annotated infection and ischaemia cases and aiming on supporting the development of DFU pathology recognition methodologies \cite{goyal2018dfunet}. DFUC 2020 was a medical imaging classification competition hosted by the Medical Image Computing and Computer Assisted Intervention (MICCAI) 2020 \cite{moi_hoon_yap_2020_3731068}. This dataset is publicly available for non-commercial research purposes only and can be obtained by emailing a formal request to the authors. The DFUC 2020 dataset consists of 4,000 images, with 2,000 used for the training set and 2,000 used for the testing set. The training set consists of DFU images only, and the testing set comprised of DFU images, other foot/skin conditions and healthy foot images. 
\par The images were captured during regular patient appointments at Lancashire Teaching Hospitals foot clinics; therefore, some images were taken from the same subjects at different intervals. Thus, the same ulcer may be present in the dataset more than once, but at different stages of development, at different angles and lighting conditions.

\subsection{Performance metrics}
Root mean square error (RMSE), peak signal-to-noise ratio (PSNR), and structural similarity index measure (SSIM) are used to evaluate both the proposed encryption and decryption algorithms and the super-resolution models quality \cite{sara2019image}. 
\par Generally, mean square error-based metrics are in common use as objective measures of distortion, due mainly to their ease of calculation. MSE value denotes the average difference of the pixels all over the image. A higher value of MSE designates a greater difference between the original image and processed image. Nonetheless, it is indispensable to be extremely careful with the edges. The following equation provides a formula for calculation of the MSE:

\begin{equation}
  MSE = \frac{1}{N}\Sigma \Sigma \left(\tilde{I}_{ij} - I_{ij}\right)^{2},
\end{equation}

where $\tilde{I}_{ij}$ denotes the feature values for a specific pixel located in i-th row, j-th column. $\tilde{I}$, $I$ correspond to reconstructed/resized image and the original one, respectively. N denotes the number of pixels.

The root mean square error (RMSE) is given by as the squared root of MSE \cite{asamoah2018measuring}:
\begin{equation}
  RMSE = \sqrt{MSE}.
\end{equation}

\par The peak signal-to-noise ratio metric is commonly used to characterize reconstructed image quality. The term PSNR is an expression for the ratio between the maximum possible value (power) of a signal and the power of distorting noise that affects the quality of its representation. Because many signals have a very wide dynamic range (ratio between the largest and smallest possible values of a changeable quantity), the PSNR is usually expressed in terms of the logarithmic decibel scale.
In this work, the assumption is made that the input and output data are a standard 2D array of data or matrix.  The dimensions of the correct image matrix and the dimensions of the degraded image matrix must be identical. The mathematical representation of the PSNR is as follows:

\begin{equation}
  PSNR = 20 \log_{10}\left(\frac{max(I)}{\sqrt{MSE}}\right).
\end{equation}

The structural similarity index (SSIM) is a perceptual metric that quantifies image quality degradation \cite{dosselmann2011comprehensive} caused by processing such as data compression or by losses in data transmission. It is a full reference metric that requires two images from the same image capture, a reference image $I$  and a processed image $\tilde{I}$. The SSIM index is defined as: 

\begin{equation}
  SSIM\left(I, \tilde{I} \right) = \left[l\left(I, \tilde{I}\right)\right]^{\alpha} 
  \cdot \left[c\left(I, \tilde{I}\right)\right]^{\beta}
  \cdot \left[s\left(I, \tilde{I}\right)\right]^{\gamma},
\end{equation}

where $\alpha>0$, $\beta>0$ and $\gamma>0$ control the relative significance of each of the three terms of the index. The luminance $l(\cdot)$, contrast $c(\cdot)$, and structural components $s(\cdot)$ of the index are defined individually as:

\begin{equation}
  l\left(I, \tilde{I}\right) = \frac{2\mu_{I}\mu_{\tilde{I}}+C_1}
                                    {\mu_{I}^2\mu_{\tilde{I}}^2 +C_1},
\end{equation}

\begin{equation}
  c\left(I, \tilde{I}\right) = \frac{2\sigma_{I}\sigma_{\tilde{I}}+C_2}
                                    {\sigma_{I}^2\sigma_{\tilde{I}}^2 +C_2},
\end{equation}

\begin{equation}
  s\left(I, \tilde{I}\right) = \frac{2\sigma_{I\tilde{I}}+C_3}
                                    {\sigma_{I} \cdot \sigma_{\tilde{I}} +C_3},
\end{equation}

where $\mu_{I}$ and $\mu_{\tilde{I}}$  represent the means of the original and coded images, respectively, $\sigma_{I}$  and $\sigma_{\tilde{I}}$  represent the standard deviations, respectively, $\sigma_{I}^2$ and $\sigma_{\tilde{I}}^2$ denote the variances, respectively, and $\sigma_{I\tilde{I}}$ is the covariance of the two images. As a means of dealing with the situations in which the denominators are close to zero, the constants C1, C2 and C3 are introduced.

\subsection{Experimental results}
In the following two section quantitative results of the implemented noise removal method, as well as the super-resolution algorithms, are presented and compared. 

\subsubsection{Noise Removal Tool}

An autoencoder model was developed, trained and evaluated on 4 separate test sets.

\begin{figure}[h]
  \centering
  \includegraphics[width=0.48\linewidth]{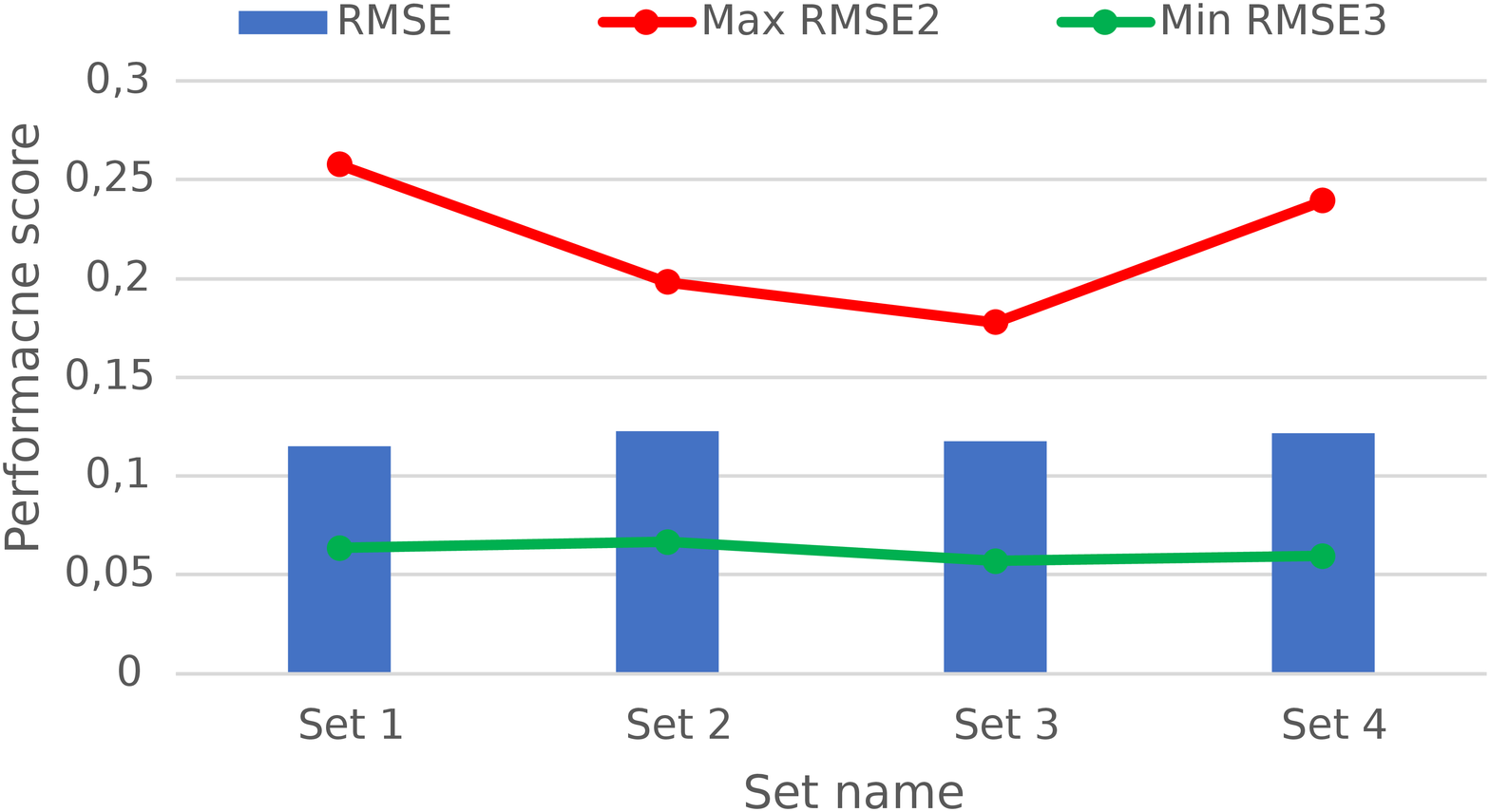}
  \includegraphics[width=0.48\linewidth]{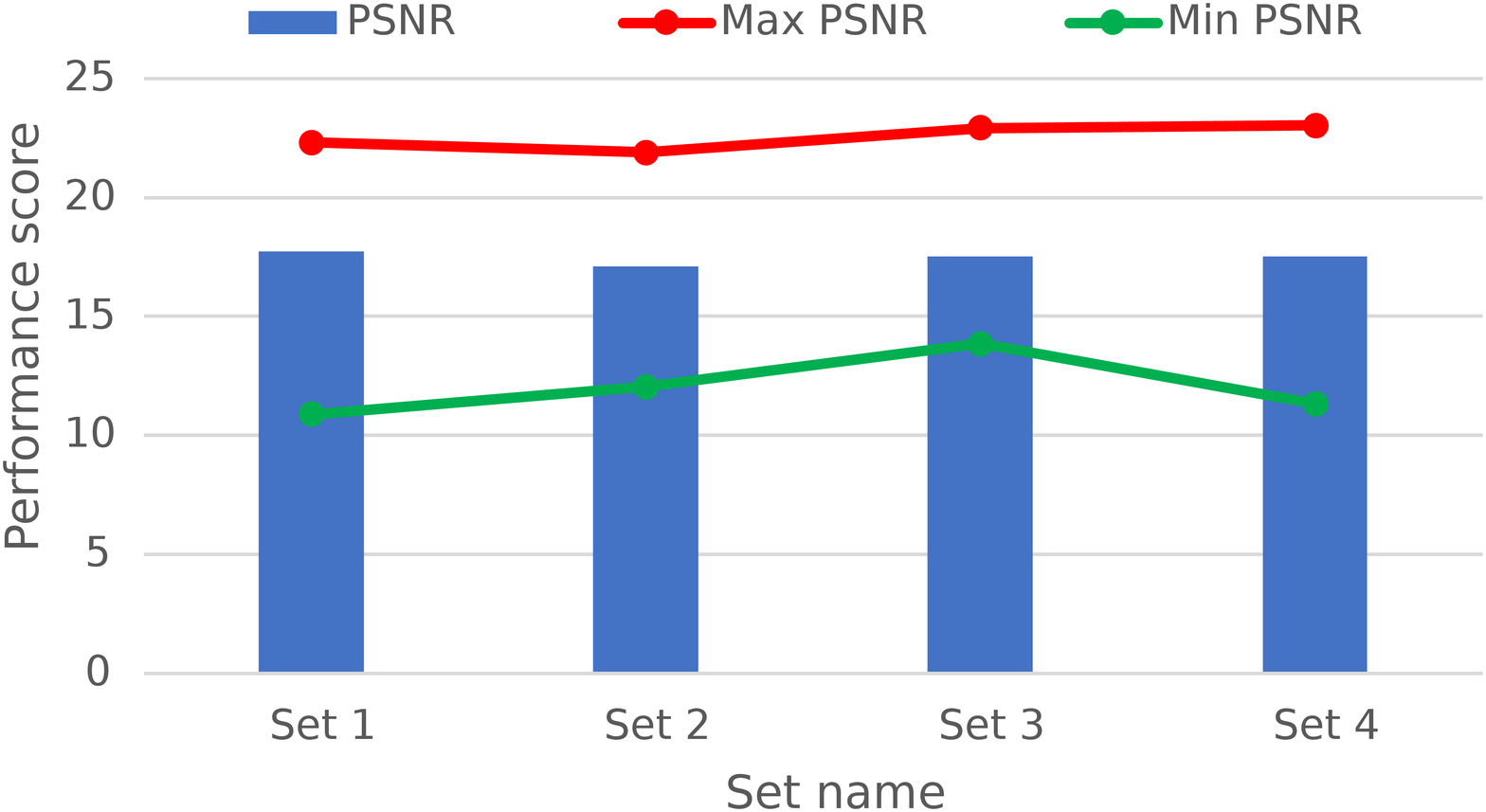}
  \caption{RMSE and PSNR scores per test set.}
  \Description{RMSE and PSNR scores per test set.}
  \label{fig:denoise_res_rmse_psnr}
\end{figure}

Figure \ref{fig:denoise_res_rmse_psnr} illustrates the RMSE and PSNR scores of the autoencoder for each test set. The model performs a RMSE of approximate 0.12 for all test sets and also a stable minimum RMSE value of around 0.07. On the other hand, the maximum RMSE value shows larger variations ranging from 0.18 to 0.27. Finally, the autoencoder performs better on set 1, followed by its evaluation on set 3. However, it should also be noted, that model’s performance on set 1 presents the largest difference between its minimum and maximum values and set 3 the smallest. The PSNR value ranges at approximately 17 for all test sets, with performance on set 3 showing the worst score and set 1 the best. The minimum PSNR values lie between 11-14, with the evaluation on set 1 having the lowest of all and on set 3 the highest. The maximum PSNR value is stable around 23 and the results on set 4 show the highest and on set 2 the lowest.

\begin{figure}[h]
  \centering
  \includegraphics[width=0.48\linewidth]{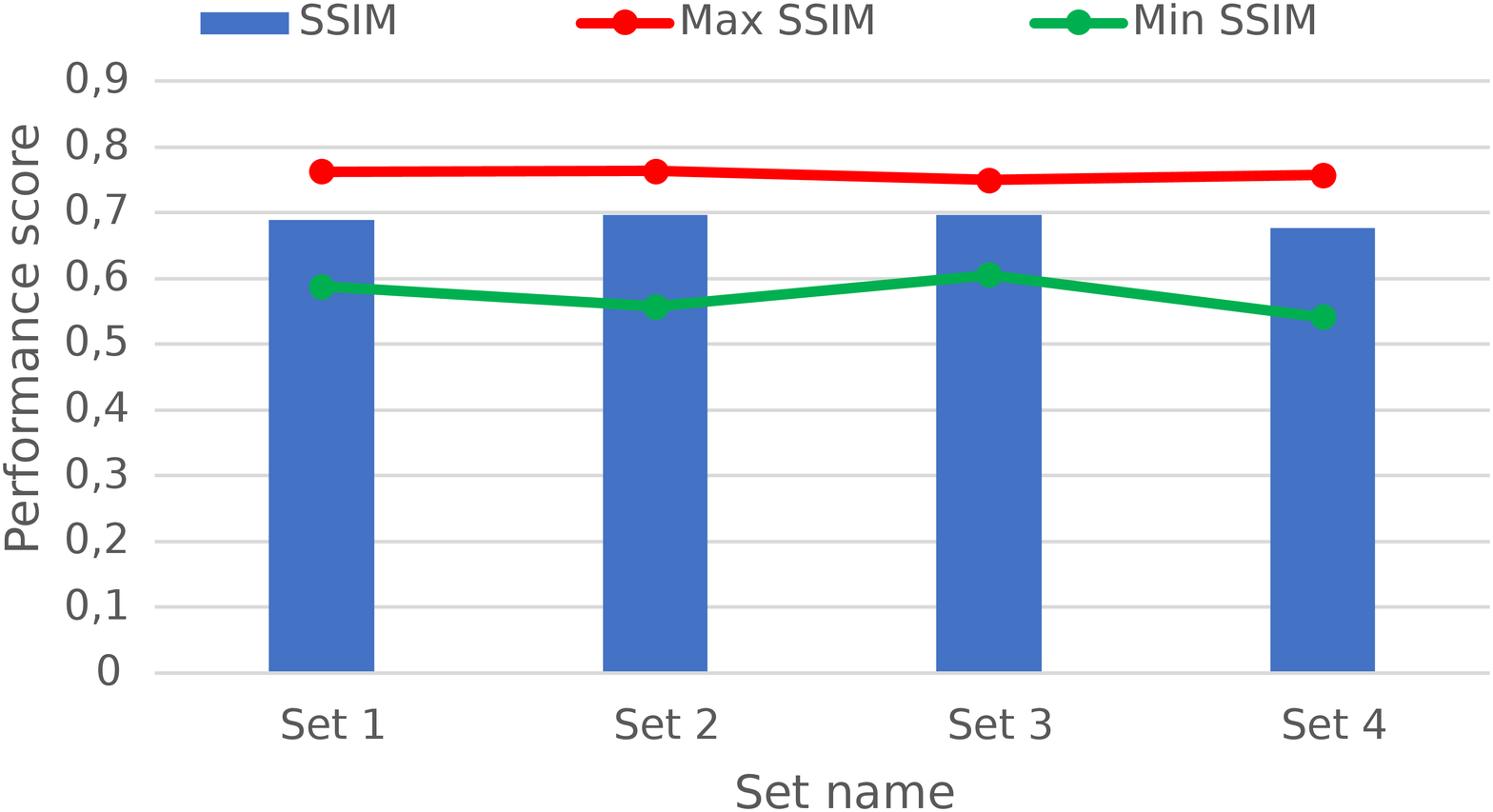}
  \includegraphics[width=0.48\linewidth]{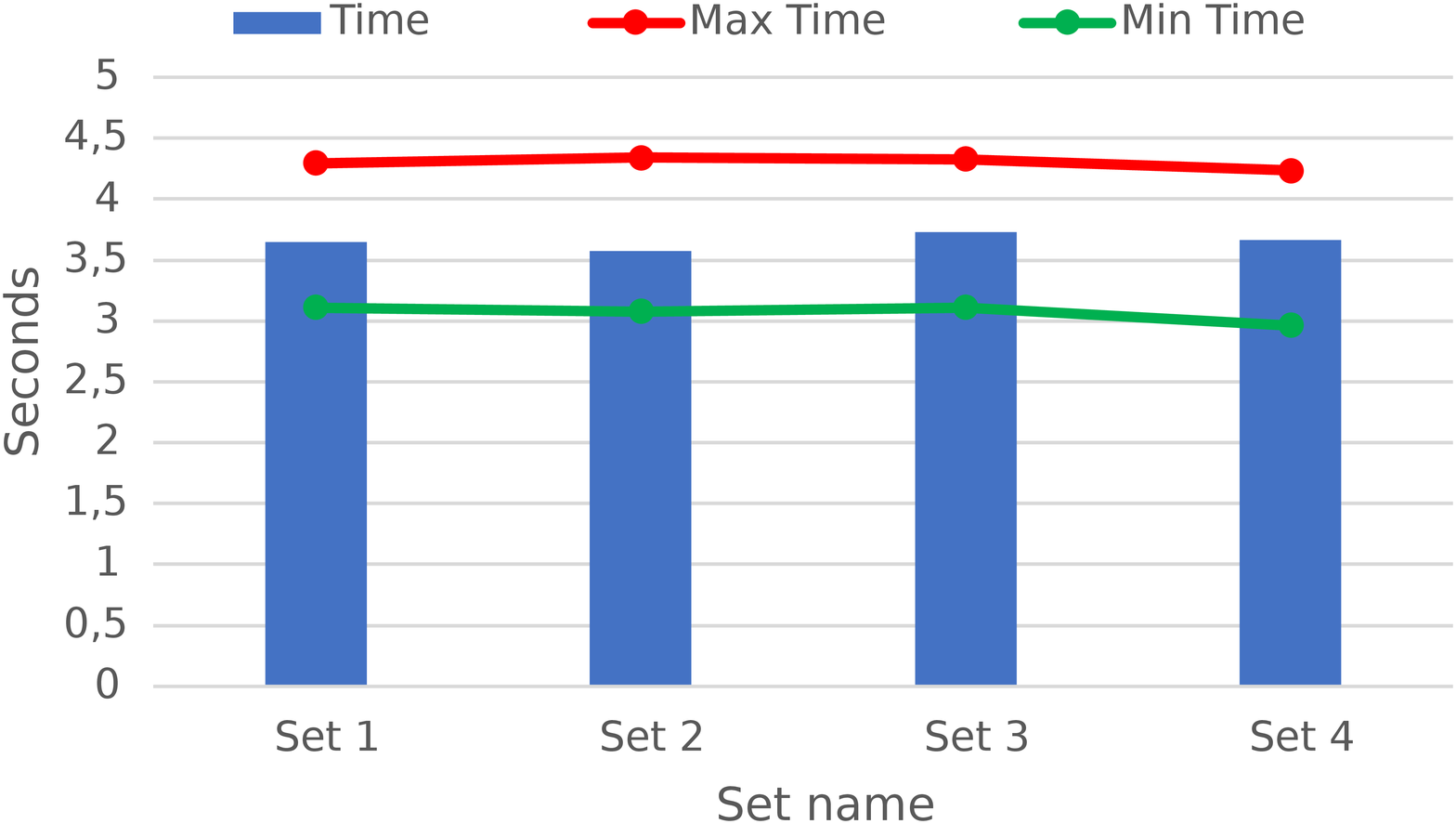}
  \caption{SSIM scores and average processing time per test set.}
  \Description{SSIM scores and average processing time per test set.}
  \label{fig:denoise_res_ssim_time}
\end{figure}

Figure \ref{fig:denoise_res_ssim_time} illustrates the scores of the autoencoder evaluated on the SSIM metric, as well as the minimum, maximum and average time the model needs to denoise a single image for each test set. The model shows a great stability with a SSIM value ranging between 0.68 and 0.7. The model performed worst on set 4 and best on sets 2 and 3 which have the same value. The minimum SSIM values lie around 0.56 and 0.6 and the maximum is almost stabilized at 0.78. In overall, the model performed best on set 3, which has the smallest difference between its minimum and maximum values, followed by the performance on set 2. The minimum and maximum values range between approximately 3 – 4.5s and the average times of the model lie between 3.5 and 3.8s. The model shows a great stability in processing time in both the average and the extreme values, as there are no large deviations when comparing the denoising times of all sets.

\begin{figure}[h]
  \centering
  \includegraphics[width=0.95\linewidth]{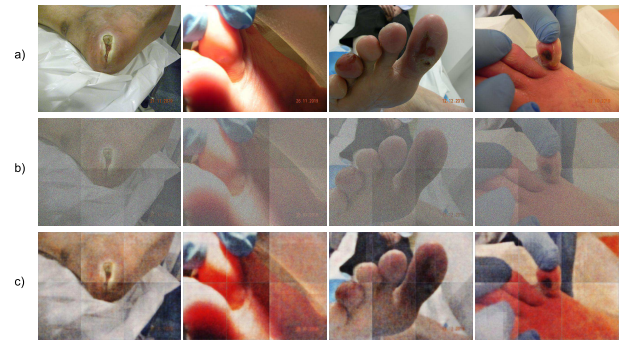}
  \caption{Demonstration of the noise remove capabilities. In (a) the original images are shown, in (b) the ones with induced noise and in (c) the resulted denoised images.}
  \Description{Demonstration of the noise remove capabilities. In column (a) the original images are shown, in (b) the ones with induced noise and in (c) the resulted denoised images.}
  \label{fig:denoise_res_images}
\end{figure}

Figure \ref{fig:denoise_res_images} provides some samples for the proposed approaches, including different camera angles. The first row shows the original images of the diabetic foot in different camera angles, the second row shows the corrupted by the noise images and the final row illustrates the denoised images after the implementation of the noise removal tools.  

\subsubsection{Super-resolution Tool}

\begin{figure}[h]
  \centering
  \includegraphics[width=0.48\linewidth]{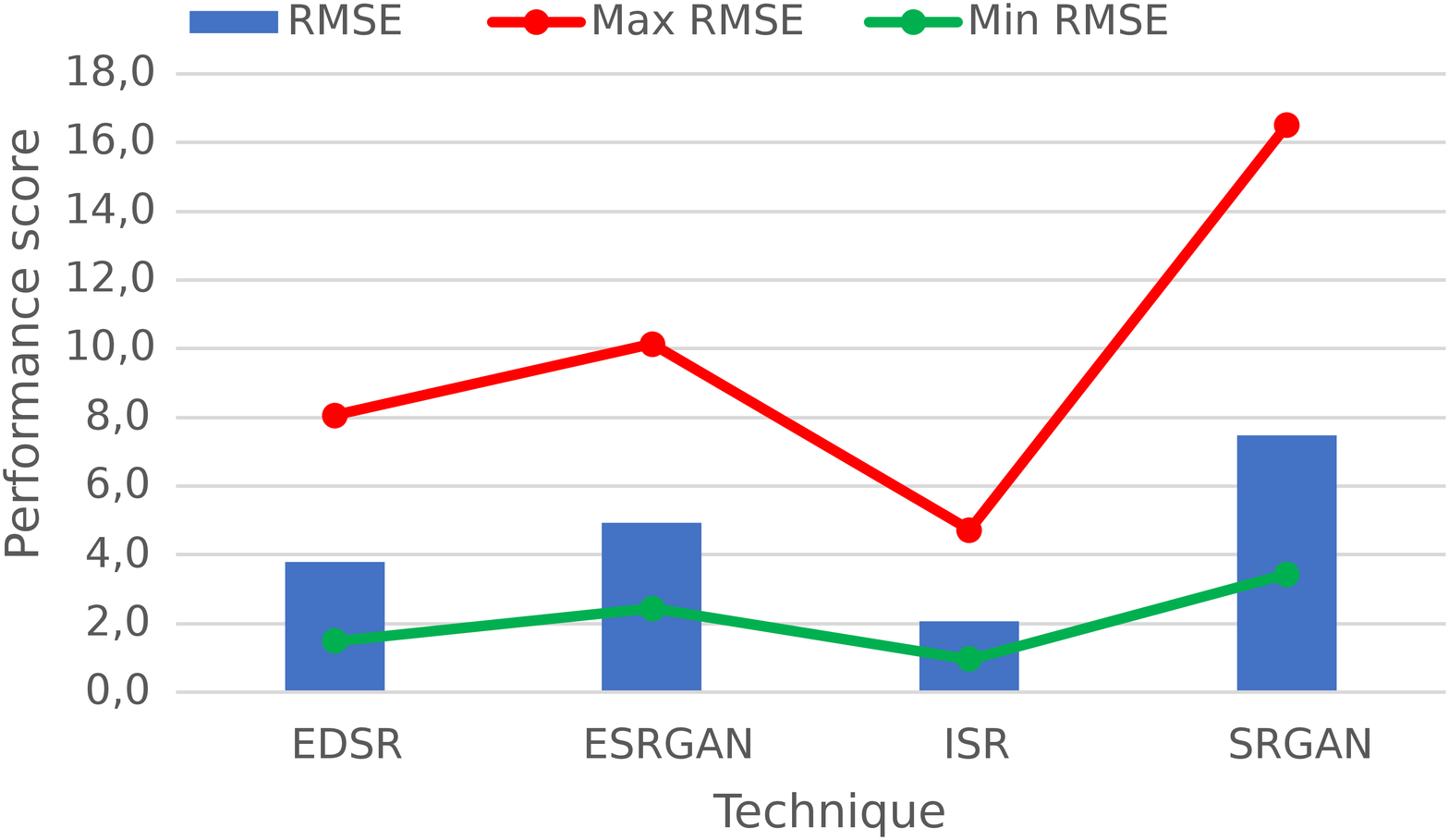}
  \includegraphics[width=0.48\linewidth]{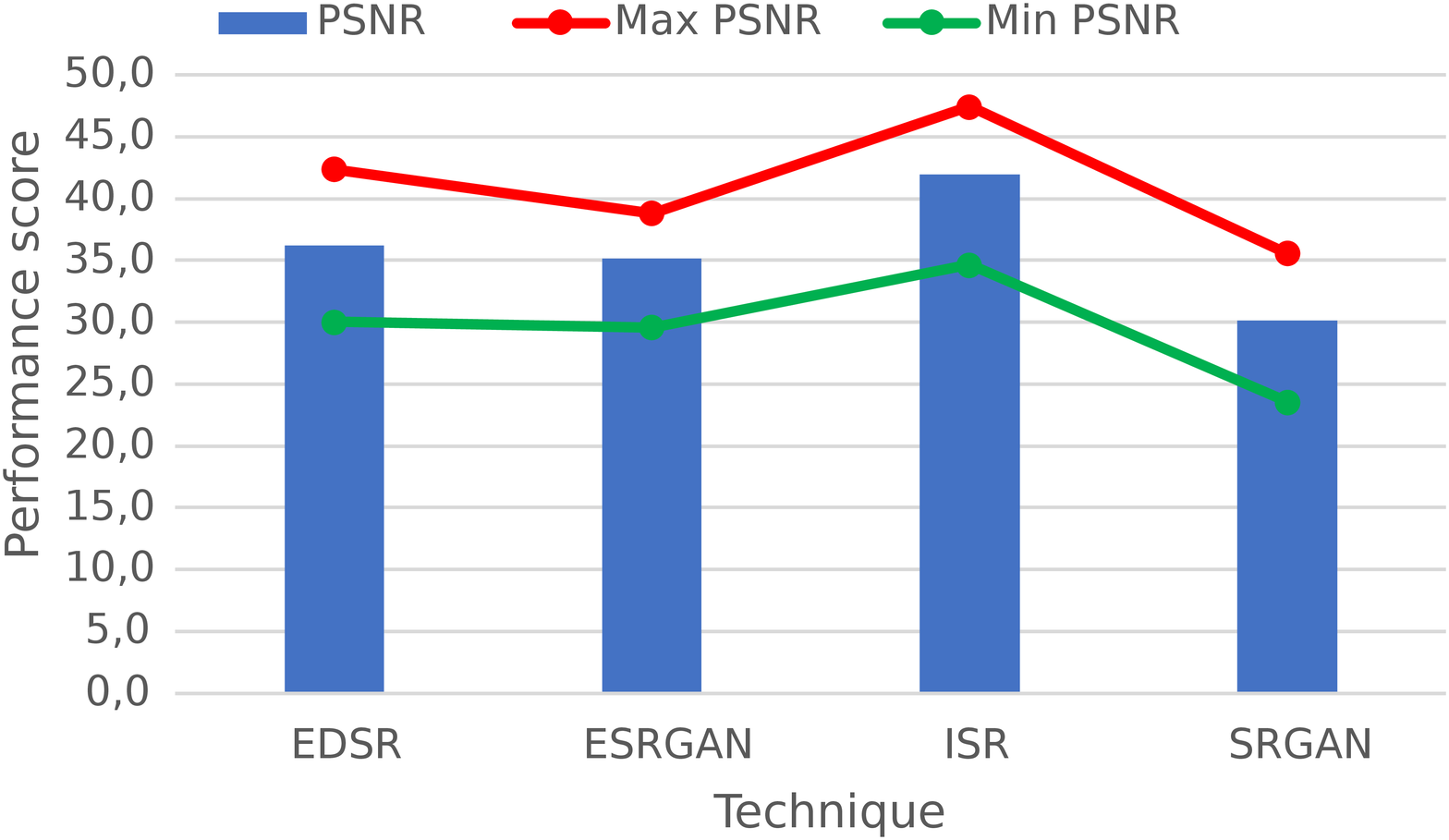}
  \caption{RMSE and PSNR scores for the adopted techniques.}
  \Description{RMSE and PSNR scores for the adopted techniques.}
  \label{fig:SR_res_rmse}
\end{figure}

In this section the four deep learning models implemented for super-resolution are compared and quantitative results are presented. Figure \ref{fig:SR_res_rmse} illustrates the RMSE and PSNR scores for the four adopted models. The ISR technique achieves the best RMSE score, with its value being value being just above 2 and its minimum and maximum values being 1 and 3, respectively. The SRGAN model on the other hand, achieves the worst RMSE score, which exceeds 7 and a maximum value above 16. The second-best technique is the EDSR having a value of almost 4, followed by the ESRGAN model with a value around 5. The best PSNR score is achieved by the ISR method with its maximum value exceeding 45 and its minimum being almost 35. The EDSR model shows the second highest PSNR with its maximum and minimum values being around 42 and 30, respectively, followed by the ESRGAN with a PSNR of 35. Finally, the SRGAN showed the worst PSNR performance, with a PSNR of 30 and its maximum value being just above 35.

\begin{figure}[h]
  \centering
  \includegraphics[width=0.48\linewidth]{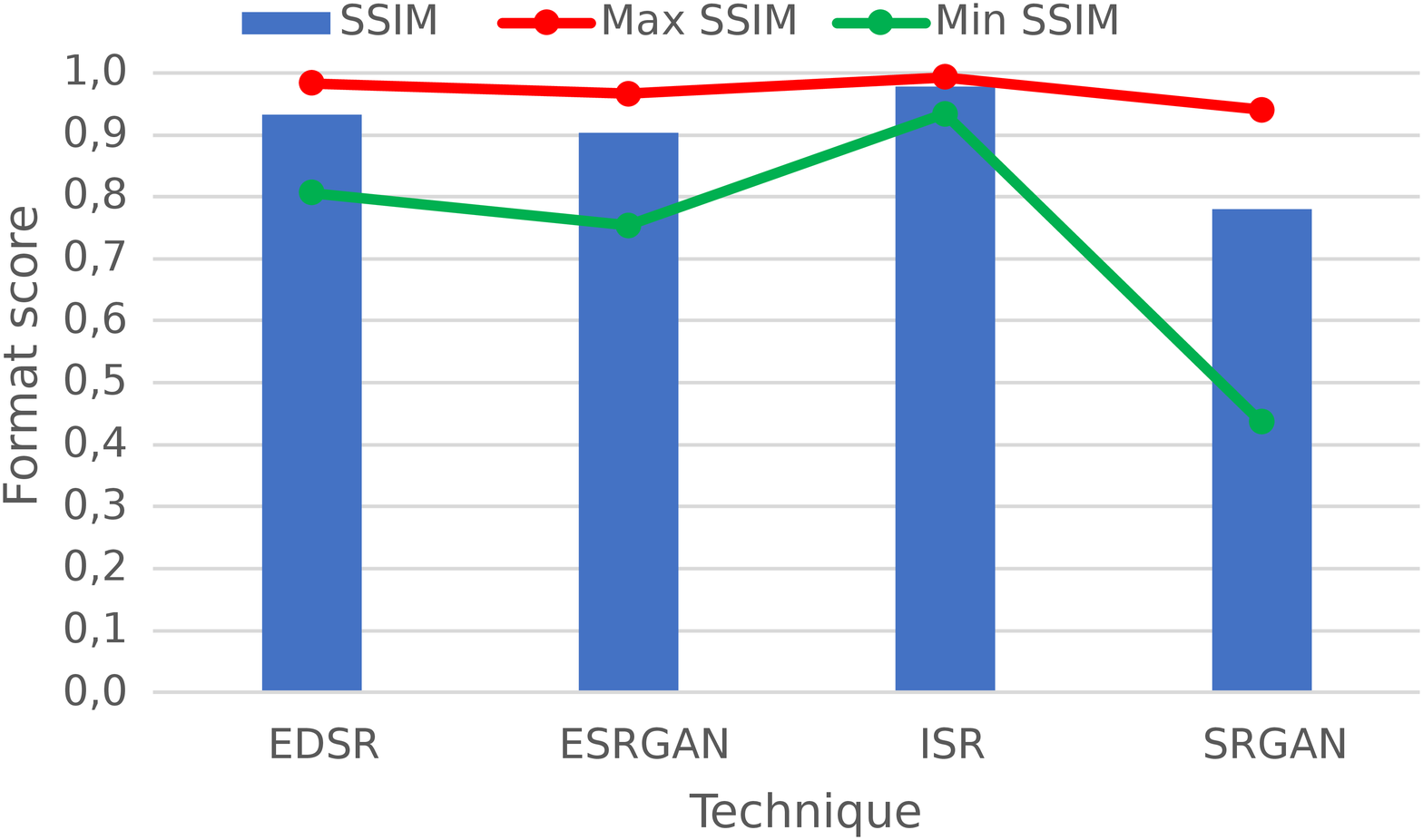}
  \includegraphics[width=0.48\linewidth]{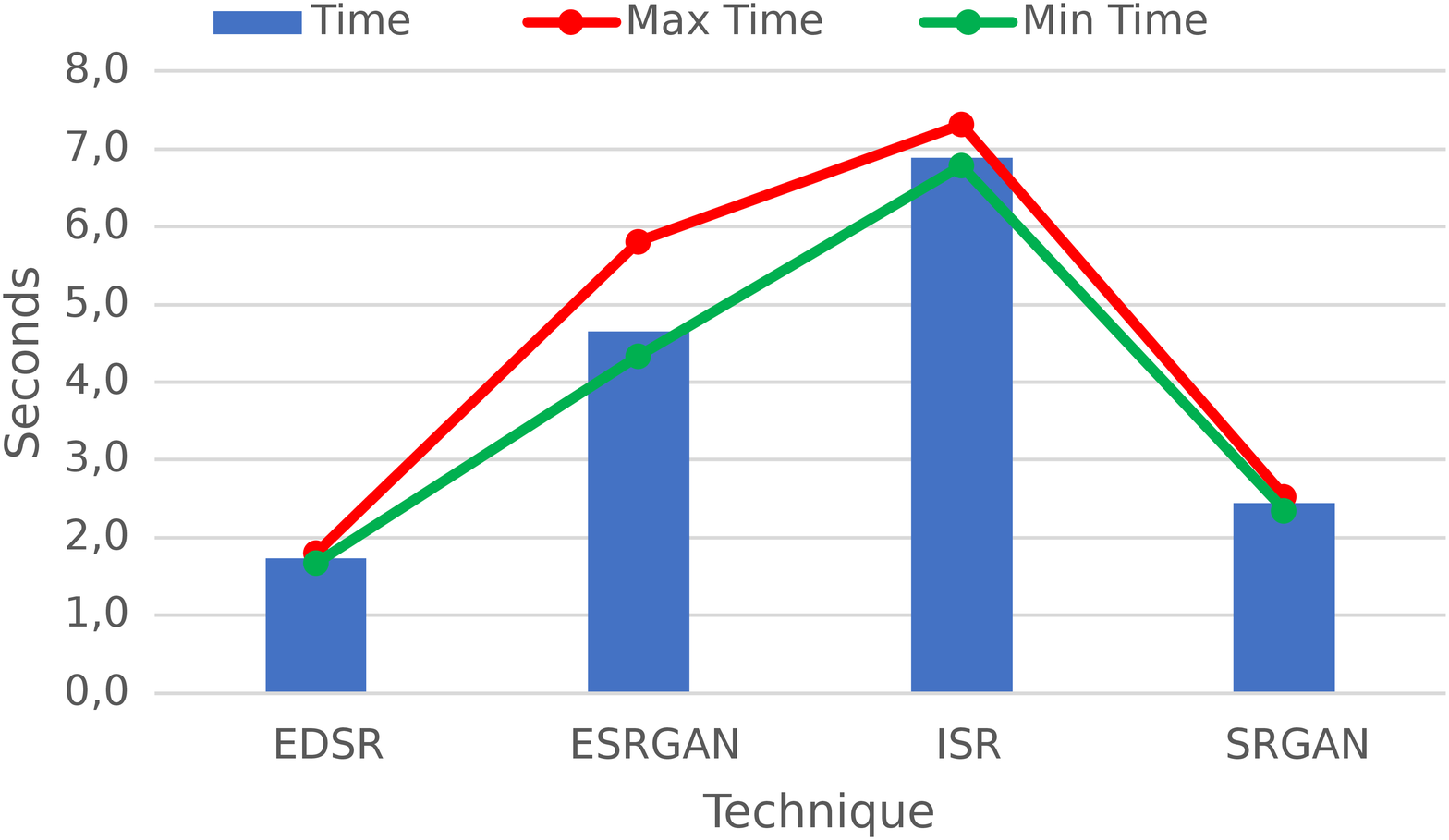}
  \caption{SSIM scores and average enhancement time, per proposed model.}
  \Description{SSIM scores and average enhancement time, per proposed model.}
  \label{fig:SR_res_ssim_time}
\end{figure}

In figure \ref{fig:SR_res_ssim_time} the super resolution methods are evaluated using the SSIM metric, as well as, the time (seconds) required to increase the resolution of an image. The results show that the ISR technique achieves the highest SSIM score which exceeds 95\% and with its minimum and maximum values lying between 93\% and 99\%. The EDSR and ESRGAN methods achieved also high SSIM scores of 93\% and 90\%, respectively. The ESRGAN shows a noticeable divergence between its minimum and maximum values. Finally, the SRGAN technique shows the worst performance with a SSIM value below 80\%. The time (seconds) required to increase the resolution of an image ranges between 1.8 and 7 seconds. Regardless of the selected approach, all the suggested schemes appear appropriate for use in hospital or home environment. Yet, the processing time is strongly depended on the available hardware.

\begin{figure}[h]
  \centering
  \includegraphics[width=0.19\linewidth]{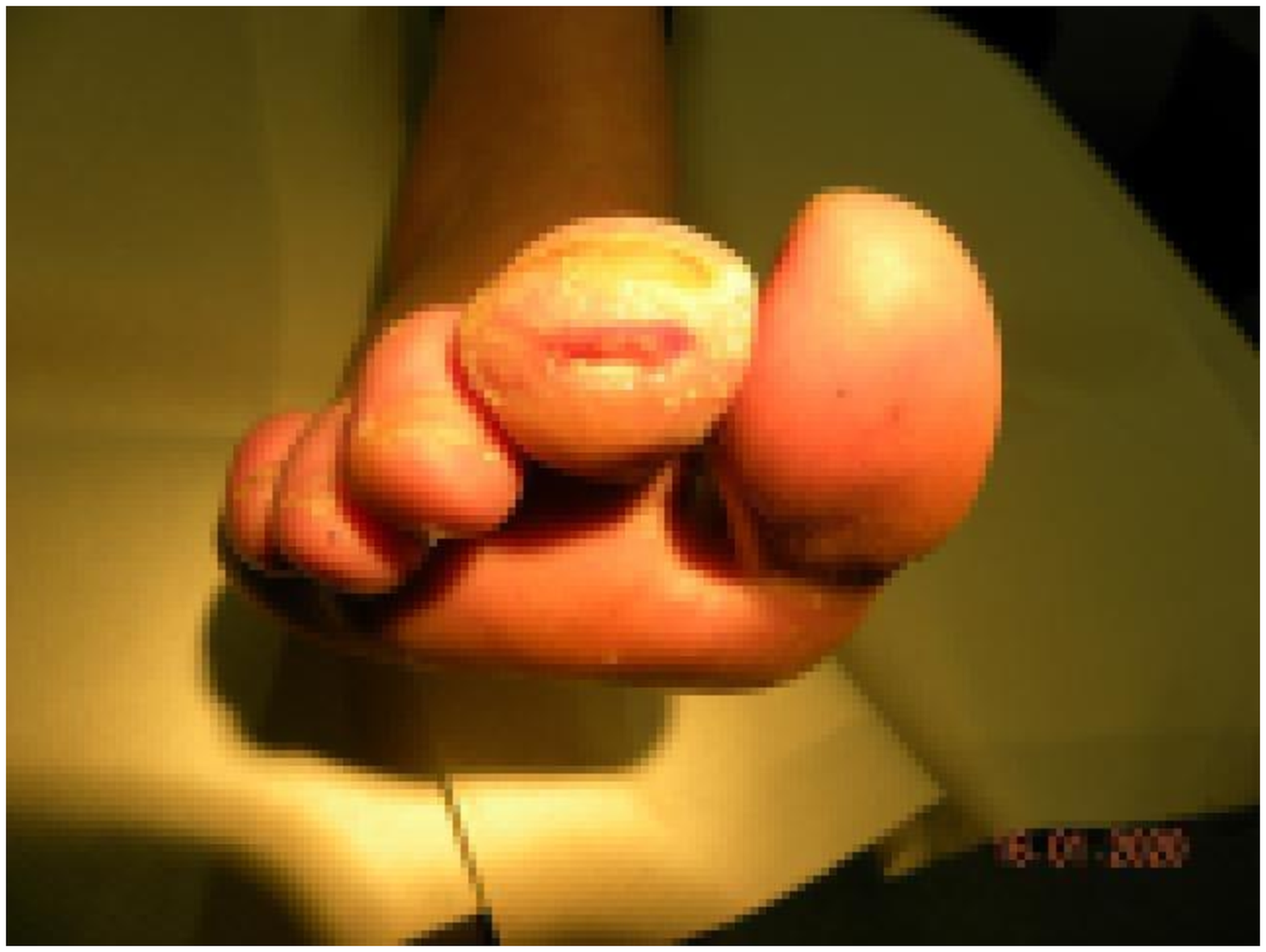} 
  \includegraphics[width=0.19\linewidth]{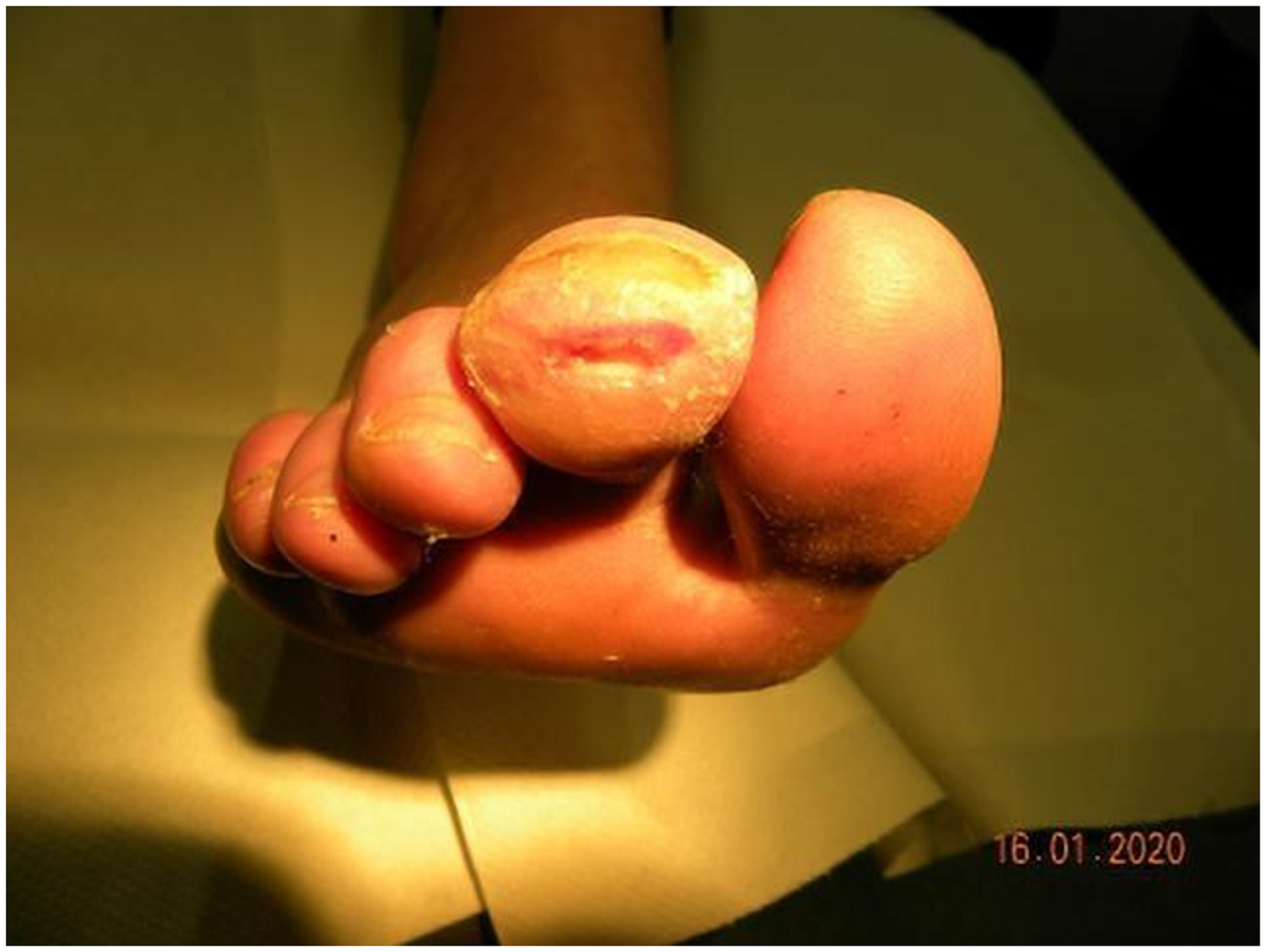}
  \includegraphics[width=0.19\linewidth]{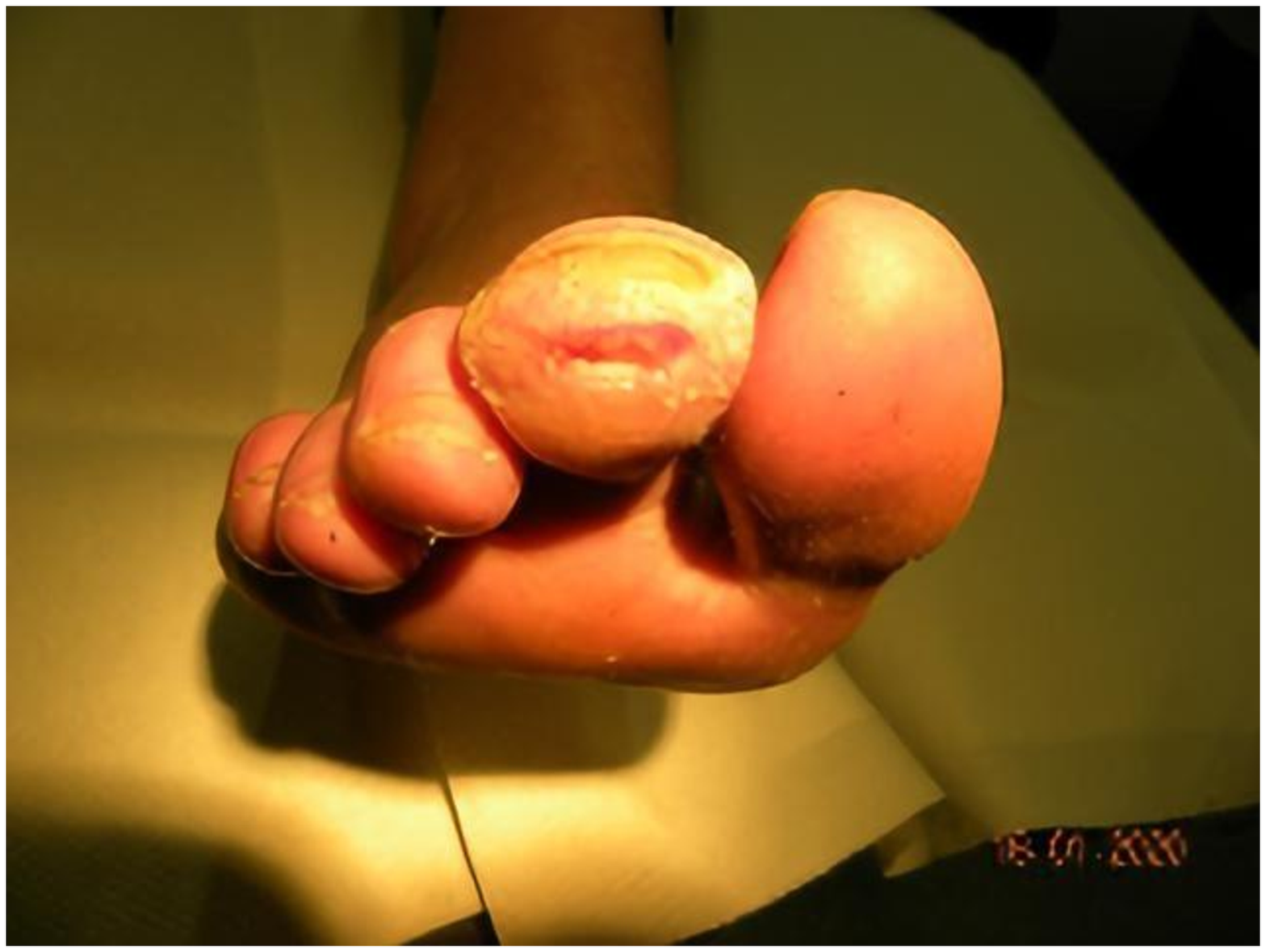}
  \includegraphics[width=0.19\linewidth]{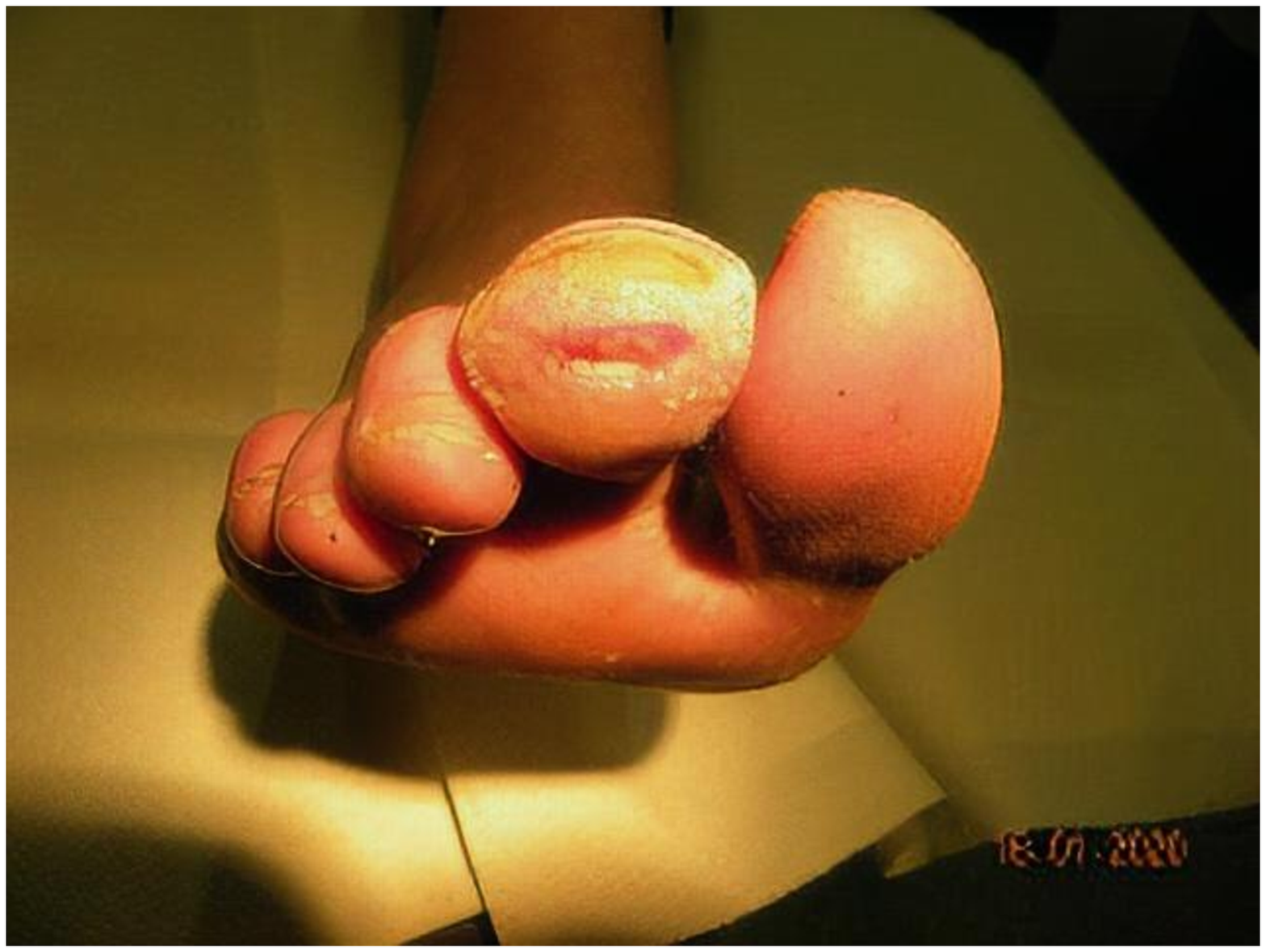}
  \includegraphics[width=0.19\linewidth]{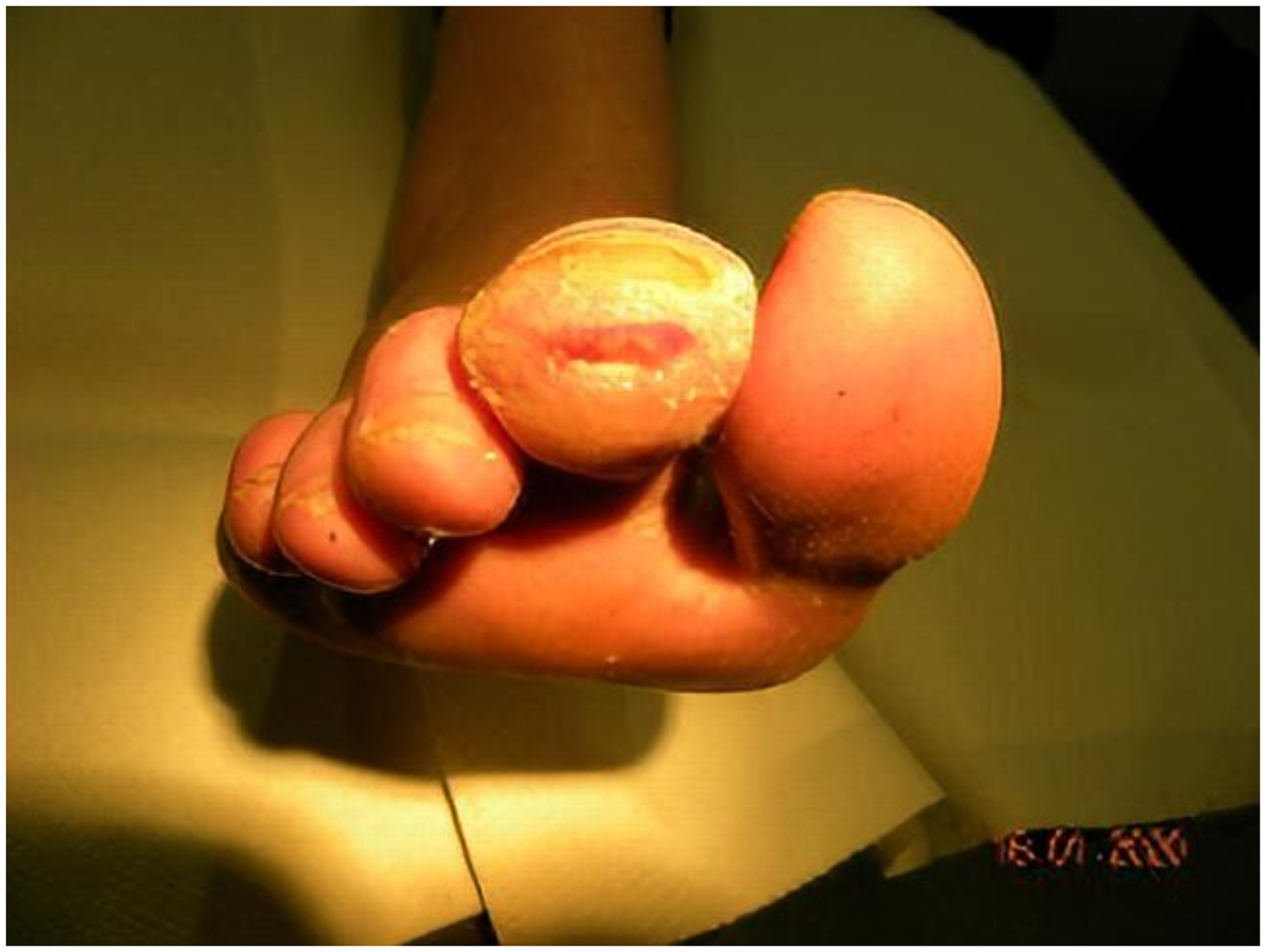}
  
  \medskip
  
  \includegraphics[width=0.19\linewidth]{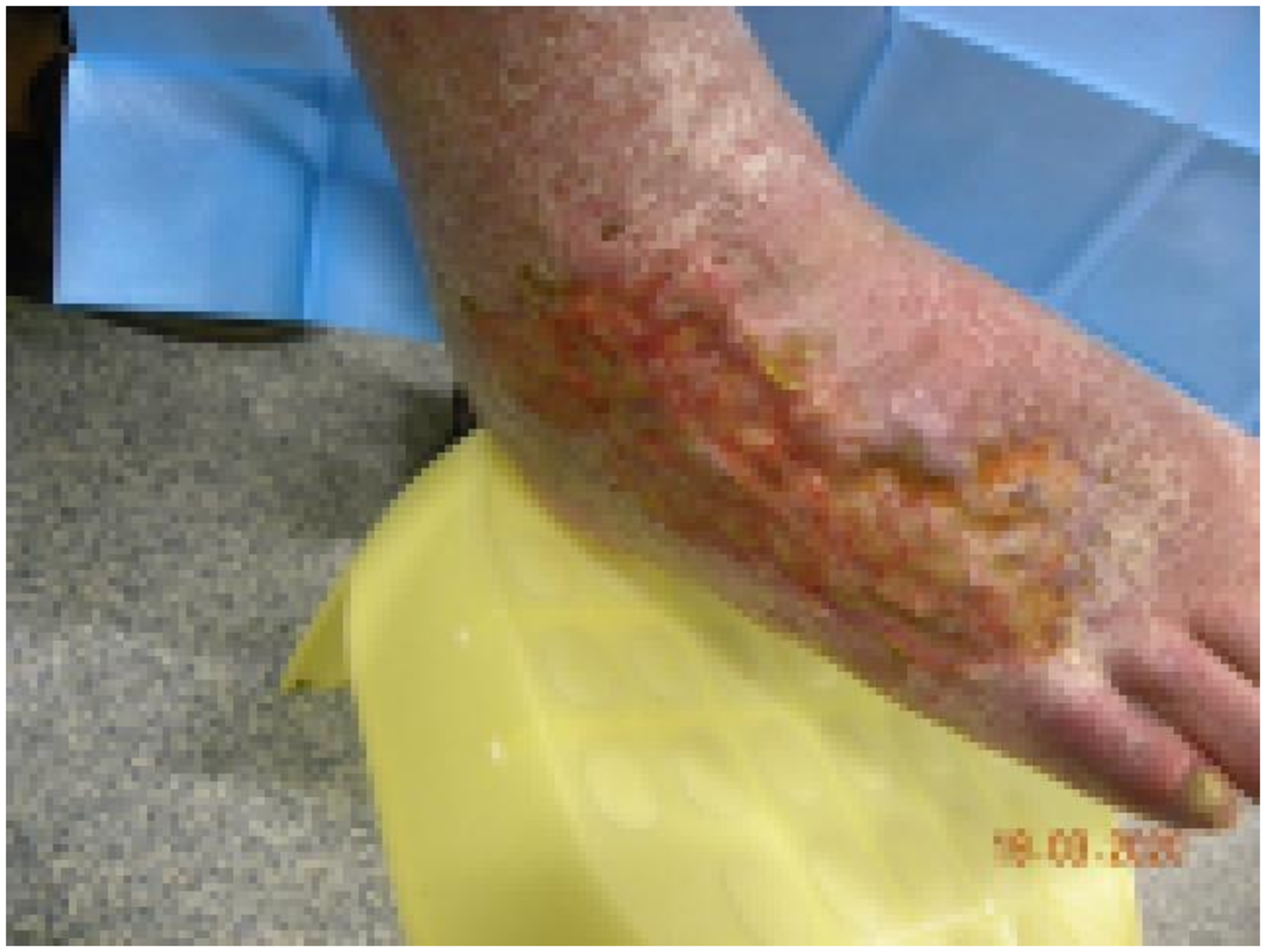} 
  \includegraphics[width=0.19\linewidth]{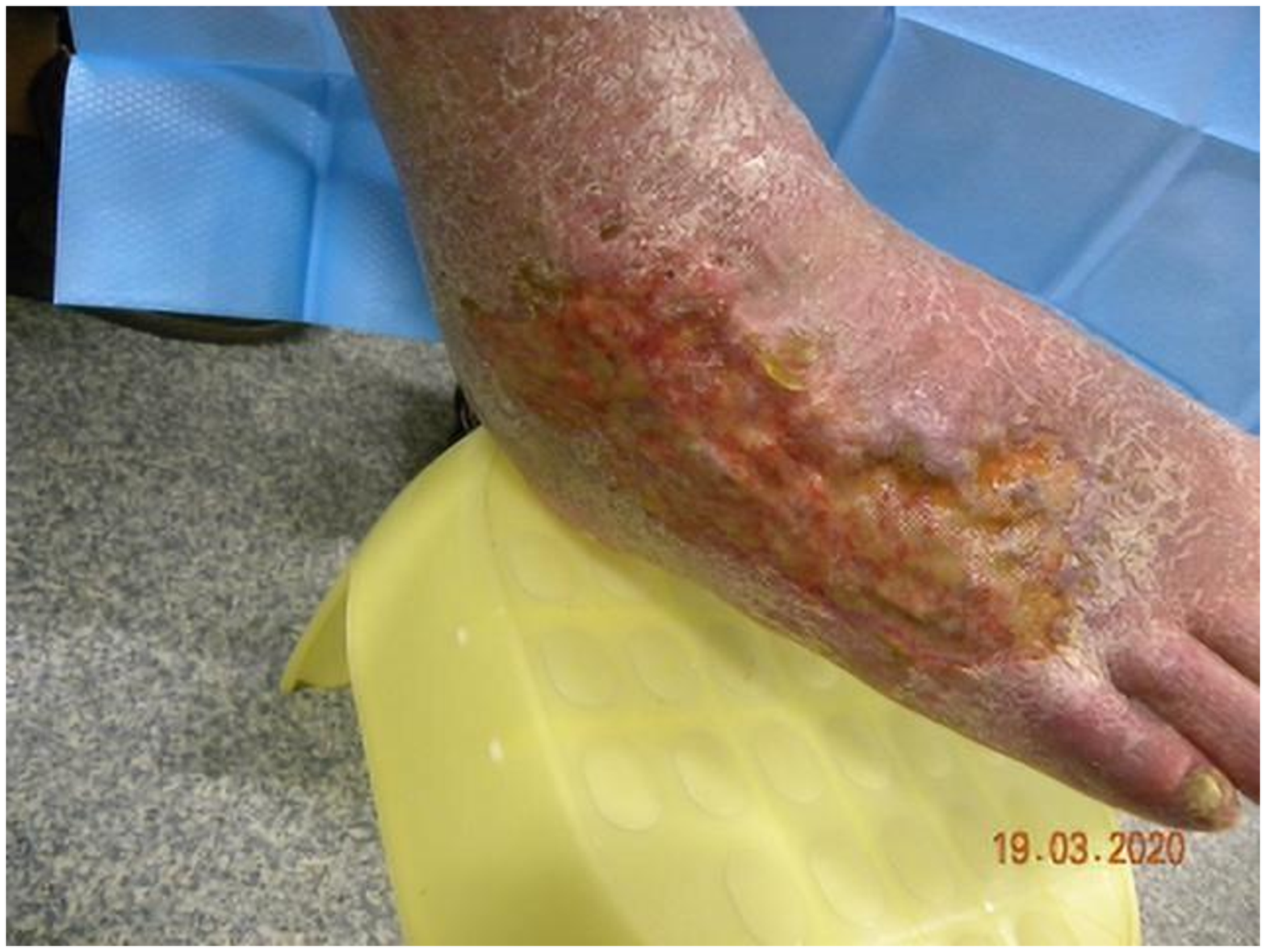} 
  \includegraphics[width=0.19\linewidth]{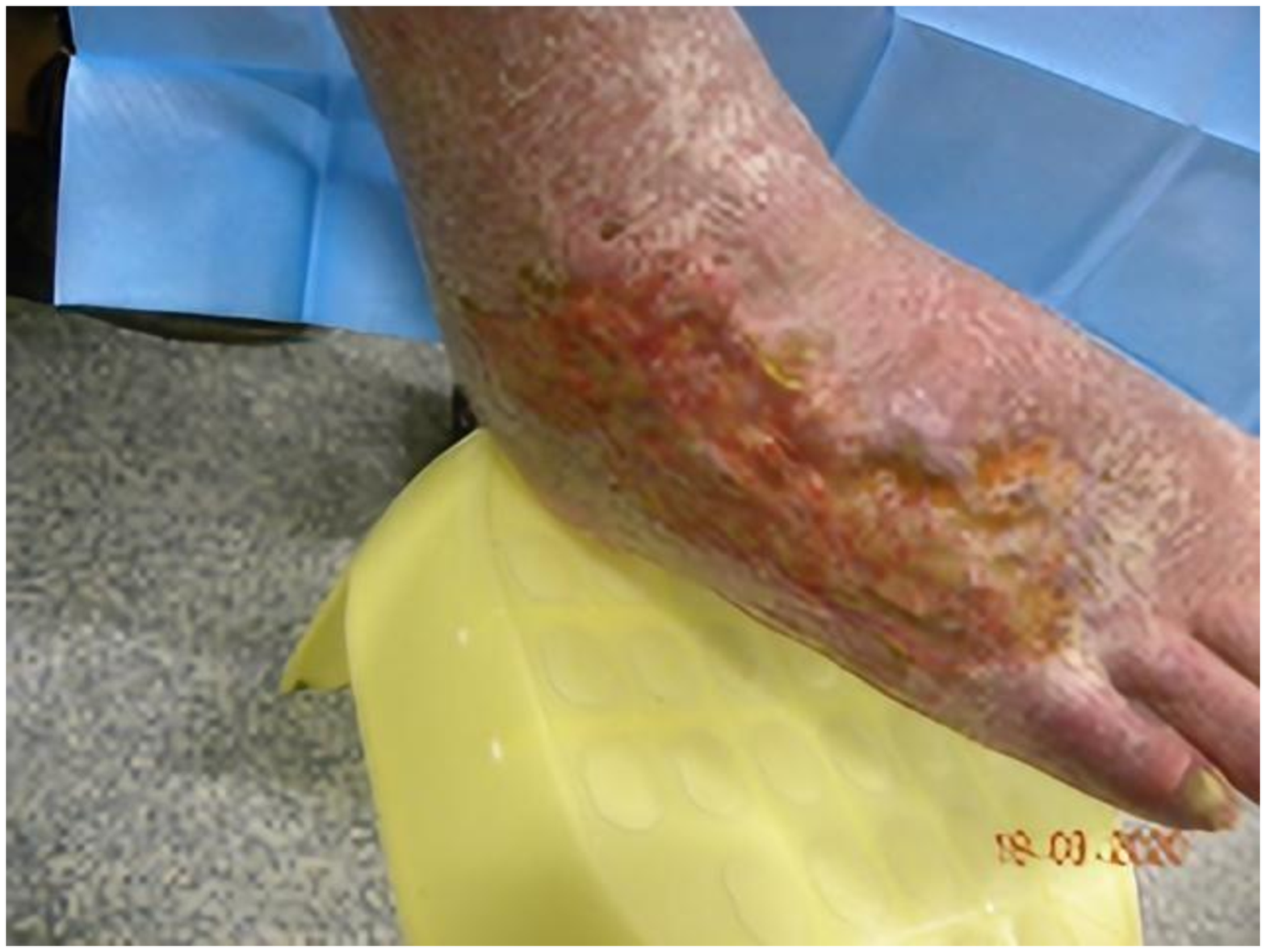} 
  \includegraphics[width=0.19\linewidth]{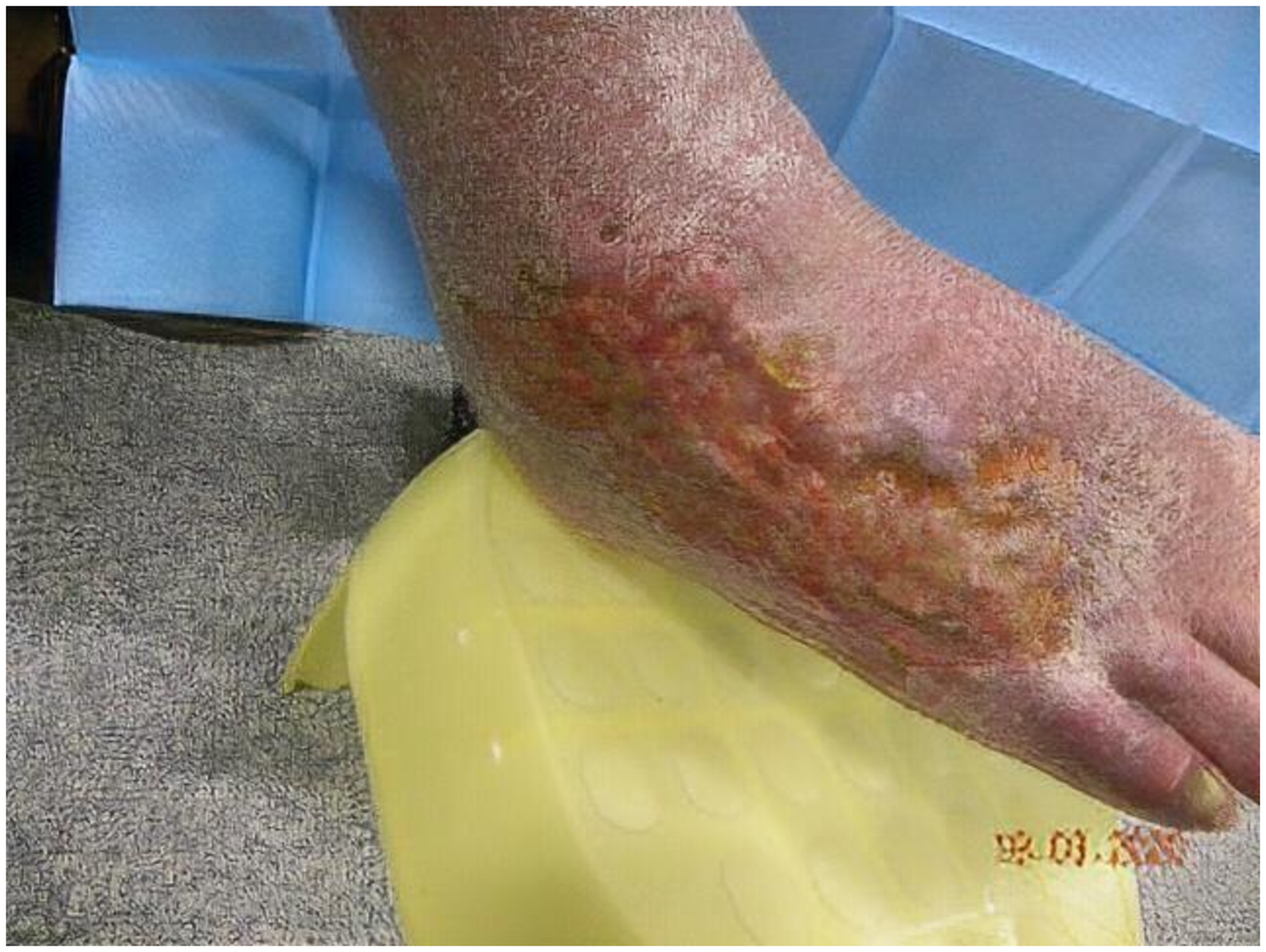} 
  \includegraphics[width=0.19\linewidth]{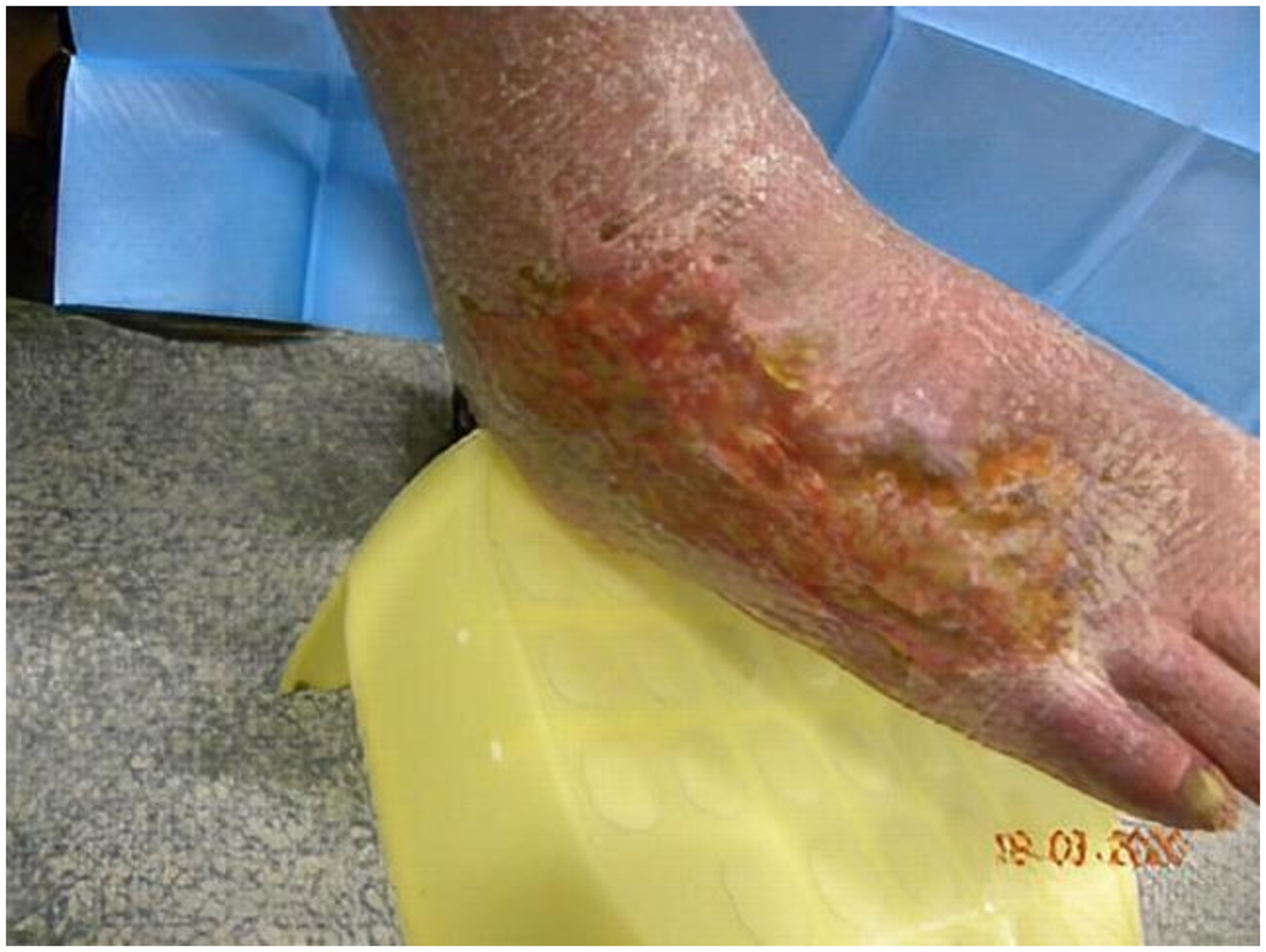}

\begin{center}
\begin{tabular}{ c c c c c }
 a \hspace{0.165\linewidth} & b \hspace{0.165\linewidth} & c \hspace{0.165\linewidth} & d \hspace{0.165\linewidth} & e  \\ 
\end{tabular}
\end{center}
  
  \caption{Comparison among Inter\_area approach (a) and the predicted images for the ISR (b), EDSR (c), SRGAN (d) and ESRGAN (e) deep learning techniques.}
  \Description{Comparison among Inter\_area approach (a) and the predicted images for the ISR (b), EDSR (c), SRGAN (d) and ESRGAN (e) deep learning techniques.}
  \label{fig:SR_res_images}
\end{figure}

Figure \ref{fig:SR_res_images} illustrates the difference in quality among different images, enhanced using suggested models. Inter\_area corresponds to a traditional linear approach and is used for demonstration purposes. ISR model performed better compared to the competition but required more processing time.

\section{Conclusion}
\label{section:concl}
In this paper we focus on image-to-image translation techniques, which can support decision making and monitoring of diabetic foot ulcers, given a set of images, provided by RGB sensors. Two cases of ItITT are considered: a) noise reduction and b) super-resolution. Starting with the noise removal tool, we have adopted an CNN-SAE model. The architecture could reconstruct the initial corrupted image despite the high standard deviation in induced noise. Then, for the resolution improvement, the ISR super-resolution tool appears to be the most prominent candidate. Processing times are less than few seconds, per image, but may vary, depending on the hardware specification of the deployed device.

%
\begin{acks}
The work in this paper has been supported by the H2020 Phootonics project: “A Cost-Effective Photonics-based Device for Early Prediction, Monitoring and Management of Diabetic Foot Ulcers” funded under the ICT H2020 framework and the grand agreement no. 871908.
\end{acks}

\bibliographystyle{ACM-Reference-Format}
\bibliography{main}


\begin{thebibliography}{38}


\ifx \showCODEN    \undefined \def \showCODEN     #1{\unskip}     \fi
\ifx \showDOI      \undefined \def \showDOI       #1{#1}\fi
\ifx \showISBNx    \undefined \def \showISBNx     #1{\unskip}     \fi
\ifx \showISBNxiii \undefined \def \showISBNxiii  #1{\unskip}     \fi
\ifx \showISSN     \undefined \def \showISSN      #1{\unskip}     \fi
\ifx \showLCCN     \undefined \def \showLCCN      #1{\unskip}     \fi
\ifx \shownote     \undefined \def \shownote      #1{#1}          \fi
\ifx \showarticletitle \undefined \def \showarticletitle #1{#1}   \fi
\ifx \showURL      \undefined \def \showURL       {\relax}        \fi
\providecommand\bibfield[2]{#2}
\providecommand\bibinfo[2]{#2}
\providecommand\natexlab[1]{#1}
\providecommand\showeprint[2][]{arXiv:#2}

\bibitem[\protect\citeauthoryear{Armstrong, Boulton, and Bus}{Armstrong
  et~al\mbox{.}}{2017}]%
        {armstrong2017diabetic}
\bibfield{author}{\bibinfo{person}{David~G Armstrong},
  \bibinfo{person}{Andrew~JM Boulton}, {and} \bibinfo{person}{Sicco~A Bus}.}
  \bibinfo{year}{2017}\natexlab{}.
\newblock \showarticletitle{Diabetic foot ulcers and their recurrence}.
\newblock \bibinfo{journal}{\emph{New England Journal of Medicine}}
  \bibinfo{volume}{376}, \bibinfo{number}{24} (\bibinfo{year}{2017}),
  \bibinfo{pages}{2367--2375}.
\newblock


\bibitem[\protect\citeauthoryear{Asamoah, Ofori, Opoku, and Danso}{Asamoah
  et~al\mbox{.}}{2018}]%
        {asamoah2018measuring}
\bibfield{author}{\bibinfo{person}{Dominic Asamoah}, \bibinfo{person}{Emmanuel
  Ofori}, \bibinfo{person}{Stephen Opoku}, {and} \bibinfo{person}{Juliana
  Danso}.} \bibinfo{year}{2018}\natexlab{}.
\newblock \showarticletitle{Measuring the performance of image contrast
  enhancement technique}.
\newblock \bibinfo{journal}{\emph{International Journal of Computer
  Applications}} \bibinfo{volume}{181}, \bibinfo{number}{22}
  (\bibinfo{year}{2018}), \bibinfo{pages}{6--13}.
\newblock


\bibitem[\protect\citeauthoryear{Bevilacqua, Roumy, Guillemot, and
  Alberi-Morel}{Bevilacqua et~al\mbox{.}}{2012}]%
        {bevilacqua2012low}
\bibfield{author}{\bibinfo{person}{Marco Bevilacqua}, \bibinfo{person}{Aline
  Roumy}, \bibinfo{person}{Christine Guillemot}, {and}
  \bibinfo{person}{Marie~Line Alberi-Morel}.} \bibinfo{year}{2012}\natexlab{}.
\newblock \showarticletitle{Low-complexity single-image super-resolution based
  on nonnegative neighbor embedding}.
\newblock  (\bibinfo{year}{2012}).
\newblock


\bibitem[\protect\citeauthoryear{Burger, Schuler, and Harmeling}{Burger
  et~al\mbox{.}}{2012}]%
        {burger2012image}
\bibfield{author}{\bibinfo{person}{Harold~C Burger},
  \bibinfo{person}{Christian~J Schuler}, {and} \bibinfo{person}{Stefan
  Harmeling}.} \bibinfo{year}{2012}\natexlab{}.
\newblock \showarticletitle{Image denoising: Can plain neural networks compete
  with BM3D?}. In \bibinfo{booktitle}{\emph{2012 IEEE conference on computer
  vision and pattern recognition}}. IEEE, \bibinfo{pages}{2392--2399}.
\newblock


\bibitem[\protect\citeauthoryear{Cassidy, Reeves, Pappachan, Gillespie,
  O’Shea, Rajbhandari, Maiya, Frank, Boulton, Armstrong,
  et~al\mbox{.}}{Cassidy et~al\mbox{.}}{2021}]%
        {cassidy2021dfuc}
\bibfield{author}{\bibinfo{person}{Bill Cassidy}, \bibinfo{person}{Neil~D
  Reeves}, \bibinfo{person}{Joseph~M Pappachan}, \bibinfo{person}{David
  Gillespie}, \bibinfo{person}{Claire O’Shea}, \bibinfo{person}{Satyan
  Rajbhandari}, \bibinfo{person}{Arun~G Maiya}, \bibinfo{person}{Eibe Frank},
  \bibinfo{person}{Andrew~JM Boulton}, \bibinfo{person}{David~G Armstrong},
  {et~al\mbox{.}}} \bibinfo{year}{2021}\natexlab{}.
\newblock \showarticletitle{The DFUC 2020 dataset: Analysis towards diabetic
  foot ulcer detection}.
\newblock \bibinfo{journal}{\emph{touchREVIEWS in Endocrinology}}
  \bibinfo{volume}{17}, \bibinfo{number}{1} (\bibinfo{year}{2021}),
  \bibinfo{pages}{5}.
\newblock


\bibitem[\protect\citeauthoryear{Dosselmann and Yang}{Dosselmann and
  Yang}{2011}]%
        {dosselmann2011comprehensive}
\bibfield{author}{\bibinfo{person}{Richard Dosselmann} {and}
  \bibinfo{person}{Xue~Dong Yang}.} \bibinfo{year}{2011}\natexlab{}.
\newblock \showarticletitle{A comprehensive assessment of the structural
  similarity index}.
\newblock \bibinfo{journal}{\emph{Signal, Image and Video Processing}}
  \bibinfo{volume}{5}, \bibinfo{number}{1} (\bibinfo{year}{2011}),
  \bibinfo{pages}{81--91}.
\newblock


\bibitem[\protect\citeauthoryear{Doulamis, Doulamis, Angeli, Lazaris, Luthman,
  Jayapala, Silbernagel, Napp, Lazarou, Karalis, et~al\mbox{.}}{Doulamis
  et~al\mbox{.}}{2021}]%
        {doulamis2021non}
\bibfield{author}{\bibinfo{person}{Anastasios Doulamis},
  \bibinfo{person}{Nikolaos Doulamis}, \bibinfo{person}{Aikaterini Angeli},
  \bibinfo{person}{Andreas Lazaris}, \bibinfo{person}{Siri Luthman},
  \bibinfo{person}{Murali Jayapala}, \bibinfo{person}{G{\"u}nther Silbernagel},
  \bibinfo{person}{Adriane Napp}, \bibinfo{person}{Ioannis Lazarou},
  \bibinfo{person}{Alexandros Karalis}, {et~al\mbox{.}}}
  \bibinfo{year}{2021}\natexlab{}.
\newblock \showarticletitle{A Non-Invasive Photonics-Based Device for
  Monitoring of Diabetic Foot Ulcers: Architectural/Sensorial Components \&
  Technical Specifications}.
\newblock \bibinfo{journal}{\emph{Inventions}} \bibinfo{volume}{6},
  \bibinfo{number}{2} (\bibinfo{year}{2021}), \bibinfo{pages}{27}.
\newblock


\bibitem[\protect\citeauthoryear{et~al.}{et~al.}{2018}]%
        {cardinale2018isr}
\bibfield{author}{\bibinfo{person}{Francesco~Cardinale et al.}}
  \bibinfo{year}{2018}\natexlab{}.
\newblock \bibinfo{title}{ISR}.
\newblock
  \bibinfo{howpublished}{\url{https://github.com/idealo/image-super-resolution}}.
\newblock


\bibitem[\protect\citeauthoryear{Ghosh, Biswas, and Ghosh}{Ghosh
  et~al\mbox{.}}{2019}]%
        {ghosh2019sdca}
\bibfield{author}{\bibinfo{person}{Swarup~Kr Ghosh}, \bibinfo{person}{Biswajit
  Biswas}, {and} \bibinfo{person}{Anupam Ghosh}.}
  \bibinfo{year}{2019}\natexlab{}.
\newblock \showarticletitle{SDCA: a novel stack deep convolutional
  autoencoder--an application on retinal image denoising}.
\newblock \bibinfo{journal}{\emph{IET Image Processing}} \bibinfo{volume}{13},
  \bibinfo{number}{14} (\bibinfo{year}{2019}), \bibinfo{pages}{2778--2789}.
\newblock


\bibitem[\protect\citeauthoryear{Gondara}{Gondara}{2016}]%
        {gondara2016medical}
\bibfield{author}{\bibinfo{person}{Lovedeep Gondara}.}
  \bibinfo{year}{2016}\natexlab{}.
\newblock \showarticletitle{Medical image denoising using convolutional
  denoising autoencoders}. In \bibinfo{booktitle}{\emph{2016 IEEE 16th
  international conference on data mining workshops (ICDMW)}}. IEEE,
  \bibinfo{pages}{241--246}.
\newblock


\bibitem[\protect\citeauthoryear{Goyal, Reeves, Davison, Rajbhandari, Spragg,
  and Yap}{Goyal et~al\mbox{.}}{2018a}]%
        {goyal2018dfunet}
\bibfield{author}{\bibinfo{person}{Manu Goyal}, \bibinfo{person}{Neil~D
  Reeves}, \bibinfo{person}{Adrian~K Davison}, \bibinfo{person}{Satyan
  Rajbhandari}, \bibinfo{person}{Jennifer Spragg}, {and}
  \bibinfo{person}{Moi~Hoon Yap}.} \bibinfo{year}{2018}\natexlab{a}.
\newblock \showarticletitle{Dfunet: Convolutional neural networks for diabetic
  foot ulcer classification}.
\newblock \bibinfo{journal}{\emph{IEEE Transactions on Emerging Topics in
  Computational Intelligence}} \bibinfo{volume}{4}, \bibinfo{number}{5}
  (\bibinfo{year}{2018}), \bibinfo{pages}{728--739}.
\newblock


\bibitem[\protect\citeauthoryear{Goyal, Reeves, Rajbhandari, Ahmad, Wang, and
  Yap}{Goyal et~al\mbox{.}}{2020}]%
        {goyal2020recognition}
\bibfield{author}{\bibinfo{person}{Manu Goyal}, \bibinfo{person}{Neil~D
  Reeves}, \bibinfo{person}{Satyan Rajbhandari}, \bibinfo{person}{Naseer
  Ahmad}, \bibinfo{person}{Chuan Wang}, {and} \bibinfo{person}{Moi~Hoon Yap}.}
  \bibinfo{year}{2020}\natexlab{}.
\newblock \showarticletitle{Recognition of ischaemia and infection in diabetic
  foot ulcers: Dataset and techniques}.
\newblock \bibinfo{journal}{\emph{Computers in biology and medicine}}
  \bibinfo{volume}{117} (\bibinfo{year}{2020}), \bibinfo{pages}{103616}.
\newblock


\bibitem[\protect\citeauthoryear{Goyal, Reeves, Rajbhandari, and Yap}{Goyal
  et~al\mbox{.}}{2018b}]%
        {goyal2018robust}
\bibfield{author}{\bibinfo{person}{Manu Goyal}, \bibinfo{person}{Neil~D
  Reeves}, \bibinfo{person}{Satyan Rajbhandari}, {and}
  \bibinfo{person}{Moi~Hoon Yap}.} \bibinfo{year}{2018}\natexlab{b}.
\newblock \showarticletitle{Robust methods for real-time diabetic foot ulcer
  detection and localization on mobile devices}.
\newblock \bibinfo{journal}{\emph{IEEE journal of biomedical and health
  informatics}} \bibinfo{volume}{23}, \bibinfo{number}{4}
  (\bibinfo{year}{2018}), \bibinfo{pages}{1730--1741}.
\newblock


\bibitem[\protect\citeauthoryear{Goyal, Yap, Reeves, Rajbhandari, and
  Spragg}{Goyal et~al\mbox{.}}{2017}]%
        {goyal2017fully}
\bibfield{author}{\bibinfo{person}{Manu Goyal}, \bibinfo{person}{Moi~Hoon Yap},
  \bibinfo{person}{Neil~D Reeves}, \bibinfo{person}{Satyan Rajbhandari}, {and}
  \bibinfo{person}{Jennifer Spragg}.} \bibinfo{year}{2017}\natexlab{}.
\newblock \showarticletitle{Fully convolutional networks for diabetic foot
  ulcer segmentation}. In \bibinfo{booktitle}{\emph{2017 IEEE international
  conference on systems, man, and cybernetics (SMC)}}. IEEE,
  \bibinfo{pages}{618--623}.
\newblock


\bibitem[\protect\citeauthoryear{He, Zhang, Ren, and Sun}{He
  et~al\mbox{.}}{2016}]%
        {he2016deep}
\bibfield{author}{\bibinfo{person}{Kaiming He}, \bibinfo{person}{Xiangyu
  Zhang}, \bibinfo{person}{Shaoqing Ren}, {and} \bibinfo{person}{Jian Sun}.}
  \bibinfo{year}{2016}\natexlab{}.
\newblock \showarticletitle{Deep residual learning for image recognition}. In
  \bibinfo{booktitle}{\emph{Proceedings of the IEEE conference on computer
  vision and pattern recognition}}. \bibinfo{pages}{770--778}.
\newblock


\bibitem[\protect\citeauthoryear{Ioffe and Szegedy}{Ioffe and Szegedy}{2015}]%
        {ioffe2015batch}
\bibfield{author}{\bibinfo{person}{Sergey Ioffe} {and}
  \bibinfo{person}{Christian Szegedy}.} \bibinfo{year}{2015}\natexlab{}.
\newblock \showarticletitle{Batch normalization: Accelerating deep network
  training by reducing internal covariate shift}. In
  \bibinfo{booktitle}{\emph{International conference on machine learning}}.
  PMLR, \bibinfo{pages}{448--456}.
\newblock


\bibitem[\protect\citeauthoryear{Johnson, Alahi, and Fei-Fei}{Johnson
  et~al\mbox{.}}{2016}]%
        {johnson2016perceptual}
\bibfield{author}{\bibinfo{person}{Justin Johnson}, \bibinfo{person}{Alexandre
  Alahi}, {and} \bibinfo{person}{Li Fei-Fei}.} \bibinfo{year}{2016}\natexlab{}.
\newblock \showarticletitle{Perceptual losses for real-time style transfer and
  super-resolution}. In \bibinfo{booktitle}{\emph{European conference on
  computer vision}}. Springer, \bibinfo{pages}{694--711}.
\newblock


\bibitem[\protect\citeauthoryear{Jolicoeur-Martineau}{Jolicoeur-Martineau}{2018}]%
        {jolicoeur2018relativistic}
\bibfield{author}{\bibinfo{person}{Alexia Jolicoeur-Martineau}.}
  \bibinfo{year}{2018}\natexlab{}.
\newblock \showarticletitle{The relativistic discriminator: a key element
  missing from standard GAN}.
\newblock \bibinfo{journal}{\emph{arXiv preprint arXiv:1807.00734}}
  (\bibinfo{year}{2018}).
\newblock


\bibitem[\protect\citeauthoryear{Kaji and Kida}{Kaji and Kida}{2019}]%
        {kaji2019overview}
\bibfield{author}{\bibinfo{person}{Shizuo Kaji} {and} \bibinfo{person}{Satoshi
  Kida}.} \bibinfo{year}{2019}\natexlab{}.
\newblock \showarticletitle{Overview of image-to-image translation by use of
  deep neural networks: denoising, super-resolution, modality conversion, and
  reconstruction in medical imaging}.
\newblock \bibinfo{journal}{\emph{Radiological physics and technology}}
  \bibinfo{volume}{12}, \bibinfo{number}{3} (\bibinfo{year}{2019}),
  \bibinfo{pages}{235--248}.
\newblock


\bibitem[\protect\citeauthoryear{Ledig, Theis, Husz{\'a}r, Caballero,
  Cunningham, Acosta, Aitken, Tejani, Totz, Wang, et~al\mbox{.}}{Ledig
  et~al\mbox{.}}{2017}]%
        {ledig2017photo}
\bibfield{author}{\bibinfo{person}{Christian Ledig}, \bibinfo{person}{Lucas
  Theis}, \bibinfo{person}{Ferenc Husz{\'a}r}, \bibinfo{person}{Jose
  Caballero}, \bibinfo{person}{Andrew Cunningham}, \bibinfo{person}{Alejandro
  Acosta}, \bibinfo{person}{Andrew Aitken}, \bibinfo{person}{Alykhan Tejani},
  \bibinfo{person}{Johannes Totz}, \bibinfo{person}{Zehan Wang},
  {et~al\mbox{.}}} \bibinfo{year}{2017}\natexlab{}.
\newblock \showarticletitle{Photo-realistic single image super-resolution using
  a generative adversarial network}. In \bibinfo{booktitle}{\emph{Proceedings
  of the IEEE conference on computer vision and pattern recognition}}.
  \bibinfo{pages}{4681--4690}.
\newblock


\bibitem[\protect\citeauthoryear{Li and Wand}{Li and Wand}{2016}]%
        {li2016combining}
\bibfield{author}{\bibinfo{person}{Chuan Li} {and} \bibinfo{person}{Michael
  Wand}.} \bibinfo{year}{2016}\natexlab{}.
\newblock \showarticletitle{Combining markov random fields and convolutional
  neural networks for image synthesis}. In
  \bibinfo{booktitle}{\emph{Proceedings of the IEEE conference on computer
  vision and pattern recognition}}. \bibinfo{pages}{2479--2486}.
\newblock


\bibitem[\protect\citeauthoryear{Lim, Son, Kim, Nah, and Mu~Lee}{Lim
  et~al\mbox{.}}{2017}]%
        {lim2017enhanced}
\bibfield{author}{\bibinfo{person}{Bee Lim}, \bibinfo{person}{Sanghyun Son},
  \bibinfo{person}{Heewon Kim}, \bibinfo{person}{Seungjun Nah}, {and}
  \bibinfo{person}{Kyoung Mu~Lee}.} \bibinfo{year}{2017}\natexlab{}.
\newblock \showarticletitle{Enhanced deep residual networks for single image
  super-resolution}. In \bibinfo{booktitle}{\emph{Proceedings of the IEEE
  conference on computer vision and pattern recognition workshops}}.
  \bibinfo{pages}{136--144}.
\newblock


\bibitem[\protect\citeauthoryear{Liu and Zhang}{Liu and Zhang}{2018}]%
        {liu2018low}
\bibfield{author}{\bibinfo{person}{Yan Liu} {and} \bibinfo{person}{Yi Zhang}.}
  \bibinfo{year}{2018}\natexlab{}.
\newblock \showarticletitle{Low-dose CT restoration via stacked sparse
  denoising autoencoders}.
\newblock \bibinfo{journal}{\emph{Neurocomputing}}  \bibinfo{volume}{284}
  (\bibinfo{year}{2018}), \bibinfo{pages}{80--89}.
\newblock


\bibitem[\protect\citeauthoryear{Oulhaj, Amine, Rziza, and Aboutajdine}{Oulhaj
  et~al\mbox{.}}{2012}]%
        {hind2012noise}
\bibfield{author}{\bibinfo{person}{Hind Oulhaj}, \bibinfo{person}{Aouatif
  Amine}, \bibinfo{person}{Mohammed Rziza}, {and} \bibinfo{person}{Driss
  Aboutajdine}.} \bibinfo{year}{2012}\natexlab{}.
\newblock \showarticletitle{Noise Reduction in Medical Images - comparison of
  noise removal algorithms -}. In \bibinfo{booktitle}{\emph{2012 International
  Conference on Multimedia Computing and Systems}}. \bibinfo{pages}{344--349}.
\newblock
\urldef\tempurl%
\url{https://doi.org/10.1109/ICMCS.2012.6320218}
\showDOI{\tempurl}


\bibitem[\protect\citeauthoryear{S{\'a}nchez and Vilaplana}{S{\'a}nchez and
  Vilaplana}{2018}]%
        {sanchez2018brain}
\bibfield{author}{\bibinfo{person}{Irina S{\'a}nchez} {and}
  \bibinfo{person}{Ver{\'o}nica Vilaplana}.} \bibinfo{year}{2018}\natexlab{}.
\newblock \showarticletitle{Brain MRI super-resolution using 3D generative
  adversarial networks}.
\newblock \bibinfo{journal}{\emph{arXiv preprint arXiv:1812.11440}}
  (\bibinfo{year}{2018}).
\newblock


\bibitem[\protect\citeauthoryear{Sara, Akter, and Uddin}{Sara
  et~al\mbox{.}}{2019}]%
        {sara2019image}
\bibfield{author}{\bibinfo{person}{Umme Sara}, \bibinfo{person}{Morium Akter},
  {and} \bibinfo{person}{Mohammad~Shorif Uddin}.}
  \bibinfo{year}{2019}\natexlab{}.
\newblock \showarticletitle{Image quality assessment through FSIM, SSIM, MSE
  and PSNR—a comparative study}.
\newblock \bibinfo{journal}{\emph{Journal of Computer and Communications}}
  \bibinfo{volume}{7}, \bibinfo{number}{3} (\bibinfo{year}{2019}),
  \bibinfo{pages}{8--18}.
\newblock


\bibitem[\protect\citeauthoryear{Timofte, De~Smet, and Van~Gool}{Timofte
  et~al\mbox{.}}{2013}]%
        {timofte2013anchored}
\bibfield{author}{\bibinfo{person}{Radu Timofte}, \bibinfo{person}{Vincent
  De~Smet}, {and} \bibinfo{person}{Luc Van~Gool}.}
  \bibinfo{year}{2013}\natexlab{}.
\newblock \showarticletitle{Anchored neighborhood regression for fast
  example-based super-resolution}. In \bibinfo{booktitle}{\emph{Proceedings of
  the IEEE international conference on computer vision}}.
  \bibinfo{pages}{1920--1927}.
\newblock


\bibitem[\protect\citeauthoryear{Tulloch, Zamani, and Akrami}{Tulloch
  et~al\mbox{.}}{2020}]%
        {tulloch2020machine}
\bibfield{author}{\bibinfo{person}{Jack Tulloch}, \bibinfo{person}{Reza
  Zamani}, {and} \bibinfo{person}{Mohammad Akrami}.}
  \bibinfo{year}{2020}\natexlab{}.
\newblock \showarticletitle{Machine learning in the prevention, diagnosis and
  management of diabetic foot ulcers: a systematic review}.
\newblock \bibinfo{journal}{\emph{IEEE Access}}  \bibinfo{volume}{8}
  (\bibinfo{year}{2020}), \bibinfo{pages}{198977--199000}.
\newblock


\bibitem[\protect\citeauthoryear{Umehara, Ota, and Ishida}{Umehara
  et~al\mbox{.}}{2017}]%
        {umehara2017super}
\bibfield{author}{\bibinfo{person}{Kensuke Umehara}, \bibinfo{person}{Junko
  Ota}, {and} \bibinfo{person}{Takayuki Ishida}.}
  \bibinfo{year}{2017}\natexlab{}.
\newblock \showarticletitle{Super-resolution imaging of mammograms based on the
  super-resolution convolutional neural network}.
\newblock \bibinfo{journal}{\emph{Open Journal of Medical Imaging}}
  \bibinfo{volume}{7}, \bibinfo{number}{4} (\bibinfo{year}{2017}),
  \bibinfo{pages}{180--195}.
\newblock


\bibitem[\protect\citeauthoryear{Umehara, Ota, and Ishida}{Umehara
  et~al\mbox{.}}{2018}]%
        {umehara2018application}
\bibfield{author}{\bibinfo{person}{Kensuke Umehara}, \bibinfo{person}{Junko
  Ota}, {and} \bibinfo{person}{Takayuki Ishida}.}
  \bibinfo{year}{2018}\natexlab{}.
\newblock \showarticletitle{Application of super-resolution convolutional
  neural network for enhancing image resolution in chest CT}.
\newblock \bibinfo{journal}{\emph{Journal of digital imaging}}
  \bibinfo{volume}{31}, \bibinfo{number}{4} (\bibinfo{year}{2018}),
  \bibinfo{pages}{441--450}.
\newblock


\bibitem[\protect\citeauthoryear{Vincent, Larochelle, Bengio, and
  Manzagol}{Vincent et~al\mbox{.}}{2008}]%
        {vincent2008extracting}
\bibfield{author}{\bibinfo{person}{Pascal Vincent}, \bibinfo{person}{Hugo
  Larochelle}, \bibinfo{person}{Yoshua Bengio}, {and}
  \bibinfo{person}{Pierre-Antoine Manzagol}.} \bibinfo{year}{2008}\natexlab{}.
\newblock \showarticletitle{Extracting and composing robust features with
  denoising autoencoders}. In \bibinfo{booktitle}{\emph{Proceedings of the 25th
  international conference on Machine learning}}. \bibinfo{pages}{1096--1103}.
\newblock


\bibitem[\protect\citeauthoryear{Voulodimos, Doulamis, Doulamis, and
  Protopapadakis}{Voulodimos et~al\mbox{.}}{2018}]%
        {voulodimos2018deep}
\bibfield{author}{\bibinfo{person}{Athanasios Voulodimos},
  \bibinfo{person}{Nikolaos Doulamis}, \bibinfo{person}{Anastasios Doulamis},
  {and} \bibinfo{person}{Eftychios Protopapadakis}.}
  \bibinfo{year}{2018}\natexlab{}.
\newblock \showarticletitle{Deep learning for computer vision: A brief review}.
\newblock \bibinfo{journal}{\emph{Computational intelligence and neuroscience}}
   \bibinfo{volume}{2018} (\bibinfo{year}{2018}).
\newblock


\bibitem[\protect\citeauthoryear{Wang, Yu, Wu, Gu, Liu, Dong, Qiao, and
  Change~Loy}{Wang et~al\mbox{.}}{2018}]%
        {wang2018esrgan}
\bibfield{author}{\bibinfo{person}{Xintao Wang}, \bibinfo{person}{Ke Yu},
  \bibinfo{person}{Shixiang Wu}, \bibinfo{person}{Jinjin Gu},
  \bibinfo{person}{Yihao Liu}, \bibinfo{person}{Chao Dong}, \bibinfo{person}{Yu
  Qiao}, {and} \bibinfo{person}{Chen Change~Loy}.}
  \bibinfo{year}{2018}\natexlab{}.
\newblock \showarticletitle{Esrgan: Enhanced super-resolution generative
  adversarial networks}. In \bibinfo{booktitle}{\emph{Proceedings of the
  European conference on computer vision (ECCV) workshops}}.
  \bibinfo{pages}{0--0}.
\newblock


\bibitem[\protect\citeauthoryear{Xie, Xu, and Chen}{Xie et~al\mbox{.}}{2012}]%
        {xie2012image}
\bibfield{author}{\bibinfo{person}{Junyuan Xie}, \bibinfo{person}{Linli Xu},
  {and} \bibinfo{person}{Enhong Chen}.} \bibinfo{year}{2012}\natexlab{}.
\newblock \showarticletitle{Image denoising and inpainting with deep neural
  networks}. In \bibinfo{booktitle}{\emph{Advances in neural information
  processing systems}}. \bibinfo{pages}{341--349}.
\newblock


\bibitem[\protect\citeauthoryear{Yap, Cassidy, Pappachan, O’Shea, Gillespie,
  and Reeves}{Yap et~al\mbox{.}}{2021a}]%
        {yap2021analysis}
\bibfield{author}{\bibinfo{person}{Moi~Hoon Yap}, \bibinfo{person}{Bill
  Cassidy}, \bibinfo{person}{Joseph~M Pappachan}, \bibinfo{person}{Claire
  O’Shea}, \bibinfo{person}{David Gillespie}, {and} \bibinfo{person}{Neil~D
  Reeves}.} \bibinfo{year}{2021}\natexlab{a}.
\newblock \showarticletitle{Analysis towards classification of infection and
  ischaemia of diabetic foot ulcers}. In \bibinfo{booktitle}{\emph{2021 IEEE
  EMBS International Conference on Biomedical and Health Informatics (BHI)}}.
  IEEE, \bibinfo{pages}{1--4}.
\newblock


\bibitem[\protect\citeauthoryear{Yap, Hachiuma, Alavi, Br{\"u}ngel, Cassidy,
  Goyal, Zhu, R{\"u}ckert, Olshansky, Huang, et~al\mbox{.}}{Yap
  et~al\mbox{.}}{2021b}]%
        {yap2021deep}
\bibfield{author}{\bibinfo{person}{Moi~Hoon Yap}, \bibinfo{person}{Ryo
  Hachiuma}, \bibinfo{person}{Azadeh Alavi}, \bibinfo{person}{Raphael
  Br{\"u}ngel}, \bibinfo{person}{Bill Cassidy}, \bibinfo{person}{Manu Goyal},
  \bibinfo{person}{Hongtao Zhu}, \bibinfo{person}{Johannes R{\"u}ckert},
  \bibinfo{person}{Moshe Olshansky}, \bibinfo{person}{Xiao Huang},
  {et~al\mbox{.}}} \bibinfo{year}{2021}\natexlab{b}.
\newblock \showarticletitle{Deep learning in diabetic foot ulcers detection: a
  comprehensive evaluation}.
\newblock \bibinfo{journal}{\emph{Computers in Biology and Medicine}}
  (\bibinfo{year}{2021}), \bibinfo{pages}{104596}.
\newblock


\bibitem[\protect\citeauthoryear{Yap, Reeves, Boulton, Rajbhandari, Armstrong,
  Maiya, Najafi, Frank, and Wu}{Yap et~al\mbox{.}}{2020}]%
        {moi_hoon_yap_2020_3731068}
\bibfield{author}{\bibinfo{person}{Moi~Hoon Yap}, \bibinfo{person}{Neil
  Reeves}, \bibinfo{person}{Andrew Boulton}, \bibinfo{person}{Satyan
  Rajbhandari}, \bibinfo{person}{David Armstrong}, \bibinfo{person}{Arun~G.
  Maiya}, \bibinfo{person}{Bijan Najafi}, \bibinfo{person}{Eibe Frank}, {and}
  \bibinfo{person}{Justina Wu}.} \bibinfo{year}{2020}\natexlab{}.
\newblock \bibinfo{title}{Diabetic Foot Ulcers Grand Challenge 2020}.
\newblock
\newblock
\urldef\tempurl%
\url{https://doi.org/10.5281/zenodo.3731068}
\showDOI{\tempurl}


\bibitem[\protect\citeauthoryear{Zhang, Tian, Kong, Zhong, and Fu}{Zhang
  et~al\mbox{.}}{2018}]%
        {zhang2018residual}
\bibfield{author}{\bibinfo{person}{Yulun Zhang}, \bibinfo{person}{Yapeng Tian},
  \bibinfo{person}{Yu Kong}, \bibinfo{person}{Bineng Zhong}, {and}
  \bibinfo{person}{Yun Fu}.} \bibinfo{year}{2018}\natexlab{}.
\newblock \showarticletitle{Residual dense network for image super-resolution}.
  In \bibinfo{booktitle}{\emph{Proceedings of the IEEE conference on computer
  vision and pattern recognition}}. \bibinfo{pages}{2472--2481}.
\newblock


\end{thebibliography}

\end{document}